\documentclass[epj]{svjour}
\usepackage{latexsym}
\usepackage{commath}
\usepackage{lmodern}        % Use an extension of the original Computer Modern font to minimize the use of bitmapped letters.
\usepackage{graphics}
\usepackage{graphicx}
\usepackage{subcaption} % Allows the use of subfigures and enables their referencing.
\usepackage{amsmath}
\usepackage{amssymb}
\usepackage{mathtools}  % Further extensions of mathematical typesetting.
\usepackage{bm} % bold math
\usepackage{bbm} % unity bold math
\usepackage[english]{babel}
\usepackage[style=nature, sorting=none, date=year, hyperref=false, doi=false]{biblatex}
\usepackage{xcolor}
\usepackage[title,titletoc]{appendix}
\usepackage{etoolbox}
\patchcmd{\appendices}{\quad}{: }{}{}

\begin{document}
%

%\footnotemark[$^*$]\footnotetext[\thefootnote]{$^*$ These authors contributed equally to this work.} \and Nikolaus Dräger\inst{2}\footnotemark[$^*$]
\newcommand*\samethanks[1][\value{footnote}]{\footnotemark[#1]}

%----------------------------------------------------------------------------------
%--------------------------------- TITLEPAGE --------------------------------------
%----------------------------------------------------------------------------------

%\title{Insert your title here\footnote{Please insert title footnote here}}

\title{Compressing the two-particle Green’s function using wavelets:}
\subtitle{Theory and application to the Hubbard atom}

\author{Emin Moghadas\inst{1}\thanks{These authors contributed equally to this work.}\fnmsep \thanks{e-mail: emin.moghadas@tuwien.ac.at (corresponding author)} \and Nikolaus Dr\"ager\inst{2}\footnotemark[1] \and Alessandro Toschi\inst{1} \and Jiawei Zang\inst{3} \and Matija Medvidovi\'c\inst{3,4} \and Dominik Kiese\inst{4} \and Andrew J. Millis\inst{3,4} \and Anirvan M.~Sengupta\inst{4,5,6} \and Sabine Andergassen\inst{2} \and Domenico Di Sante\inst{7}} % etc

%%\offprints{}          % Insert a name or remove this line
%
\institute{Institute for Solid State Physics, TU Wien, 1040 Vienna, Austria \and Institute for Solid State Physics and Institute of Information Systems Engineering, TU Wien, 1040 Vienna, Austria \and Department of Physics, Columbia University, 538 W 120th Street, New York, New York 10027, USA \and Center for Computational Quantum Physics, Flatiron Institute, 162 5th Avenue, New York, NY 10010, USA \and Center for Computational Mathematics, Flatiron Institute, 162 5th Avenue, New York, NY 10010, USA \and Department of Physics and Astronomy, Rutgers University, Piscataway, New Jersey 08854, USA \and Department of Physics and Astronomy, University of Bologna, 40127 Bologna, Italy}
\date{Received: ... / Revised version: ...}
\abstract{Precise algorithms capable of providing controlled solutions in the presence of strong interactions are transforming the landscape of quantum many-body physics. Particularly exciting breakthroughs are enabling the computation of non-zero temperature correlation functions. However, computational challenges arise due to constraints in resources and memory limitations, especially in scenarios involving complex Green's functions and lattice effects. Leveraging the principles of signal processing and data compression, this paper explores the wavelet decomposition as a versatile and efficient method for obtaining compact and resource-efficient representations of the many-body theory of interacting systems. The effectiveness of the wavelet decomposition is illustrated through its application to the representation of generalized susceptibilities and self-energies in a prototypical interacting fermionic system, namely the Hubbard model at half-filling in its atomic limit. These results are the first proof-of-principle application of the wavelet compression within the realm of many-body physics and demonstrate the potential of this wavelet-based compression scheme for  understanding the physics of correlated electron systems.}
%
%\PACS{
%      {PACS-key}{discribing text of that key}   \and
%      {PACS-key}{discribing text of that key}
%     } % end of PACS codes
%} %end of abstract
%
%
%
%

\maketitle

%----------------------------------------------------------------------------------
%-------------------------------- INTRO -------------------------------------------
%----------------------------------------------------------------------------------

\section{Introduction}
\label{intro}

The emergence of highly precise algorithms capable of providing controlled solutions even within the realm of strong interactions is transforming the field of quantum many-body physics \cite{georges1996dynamical, maier2005, gull2011, metzner2012, rohringer2018, kugler2021, lee2021}. In various practical scenarios though, memory limitations and constraints on computational resources make achieving the necessary precision challenging. Particular challenges include the storage problems associated with two particle Green functions ~\cite{shinaoka2018overcomplete}, which depend on three independent frequencies and three independent momenta, and even for single-particle Green's functions when many degrees of freedom are involved, such as in the presence of spin-orbit coupling or in multi-orbitals applications, as often encountered, for instance, in ab-initio many-body calculations~\cite{shinaoka2022efficient}. The large size of these quantities also imposes constraints on the ability to manipulate and operate on these quantities. The Matsubara formalism commonly used for thermodynamic properties of systems at non-zero temperature ~\cite{abrikosov2012methods} leads to a \emph{polynomial} growth in size with respect to inverse temperature $\beta$, rapidly surpassing the capabilities of even state-of-the-art high-performance computing architectures. These challenges underscore the need for developing efficient compression schemes that allow, on the one hand, to conveniently store the many-body quantities, while still permitting, on the other hand, to perform operations typical to many-body physics.

By taking advantage of the specific mathematical structure of the imaginary time Green's function $G(\tau)$ given by the spectral Lehmann representation, two novel representation strategies leverage the fact that the analytic continuation kernel or Lehmann kernel can be approximated to high accuracy by a low rank decomposition to derive compressed representations of $G(\tau)$. The first approach, known as intermediate representation (IR)~\cite{shinaoka2017compressing,chikano2018performance}, uses the singular value decomposition (SVD) of the Lehmann kernel to construct orthogonal bases for the imaginary time Green's function. The IR has proven effective in numerous applications, demonstrating success also in scenarios that involve two-particle quantities~\cite{shinaoka2018overcomplete,shinaoka2020sparse,nagai2019smooth,otsuki2020sparse,wang2020efficient,shinaoka2021sparse,wallerberger2021solving}. The second compression method, which is called the discrete Lehmann representation (DLR), expands the imaginary time Green's function onto a basis set composed of exponentials, such that the corresponding expansion can be conceptualized as a discrete version of the Lehmann representation, utilizing an effective spectral density composed of a sum of $\delta$ functions~\cite{kaye2022discrete,kaye2022libdlr}. Since the IR and DLR schemes are based on a low rank representation of the spectral Lehmann kernel, they are generally restricted to the imaginary time or Matsubara domains \cite{kaye2022discrete, kaye2022libdlr}. Here we aim at a compression method that is \emph{agnostic} to the underlying representation of the data, but is still able to yield an effective and compact representation. In this context, for the treatment of quantum-field theory correlation functions, the quantics tensor train (QTT) ansatz \cite{shinaoka2023qtt} has been recently proposed. At the same time, the discrete wavelet transform (DWT) emerges as a complementary candidate, as its versatility allows to find compact representations of not only imaginary time or Matsubara data but also quantities defined for real times or frequencies as well as momentum dependent objects.

The wavelet decomposition is a powerful mathematical technique that has gained significant attention within the framework of signal processing and data compression ~\cite{haar1910,grossman1984,mallat1989theory,daubechies1990,daubechies1992ten,cohen1992,meyer1993,mallat1999wavelet}. Unlike conventional methods such as the Fourier transform, the wavelet decomposition is able to resolve both frequency and location information in a signal, making it particularly suitable at representing complex and dynamic data, while efficiently concentrating information in a few significant coefficients, thus allowing for the removal of less essential details. This adaptability enables wavelets to compress data with minimal loss of important features, and has made them one of the dominant methods for various applications, from image and audio compression to the analysis of time-series data~\cite{beylkin1991,donoho1995,donoho1995vag,abramovich1998,mallat1999wavelet,candes2000,chirstopoulos2000,sifuzzaman2009application,akansu2010emerging, abbott2016}. In this paper, we show the versatility and efficiency of the wavelet compression to address the ever-growing demand for more compact and resource-efficient data representations in the many-body theory of interacting systems.

To prove the efficiency of the wavelet decomposition in producing compact representations of many-body quantities, such as generalized susceptibilities and the self-energy of an interacting fermionic system, we focus our attention on a prototypical correlated platform: the Hubbard model at half-filling in its atomic limit (referred to as Hubbard atom in the following)~\cite{esterling1970hubbard}. Despite its very simple form, that is reflected in the possibility of generating data from closed analytical expressions at negligible computational cost~\cite{thunstrom2018analytical}, this model exhibits complex dynamical two-particle correlations, making it the perfect playground to test the validity and effectiveness of our wavelet-based compression scheme.

The paper is structured as follows: in Sec.~\ref{sec:model}, we first provide an introduction about the most important features of the Hubbard atom and subsequently in Sec.~\ref{sec:wavelet}, a primer on the mathematical basics of the wavelet transform, detailing how it can be utilized to implement a compression method through the pruning of a fraction of the smallest coefficients in terms of magnitude.
We also present and explore two metrics, the compression ratio in memory and the Structural Similarity Index Measure (SSIM), demonstrating their utility in measuring the strength of the compression and giving an introduction into the mathematical foundation behind the SSIM.
Finally, in Sec.~\ref{sec:results}, we will apply the method to the generalized spin and charge susceptibilities as well as the self-energy of the Hubbard atom.
In particular, we show that even when pruning a very large fraction of the detail coefficients, the original data can be reconstructed very accurately. 
This result suggests that only a few features in the underlying data may be critical for describing the physical phenomena of correlated Hubbard atoms. Although specific to this toy model, our work acts as a foundation for further research in this direction. While in fact the focus of this study has been predominantly the methodological development, future work may concentrate on the application of our method for a more in-depth analysis of the underlying physics of correlated electron systems beyond the atomic limit, where the competition of kinetic and potential energies generates a whole richer set of physical phenomena.

%--------------------------------------------------------------------------------------------------------
%----------------------------------------- MODELS / METHODS ---------------------------------------------
%--------------------------------------------------------------------------------------------------------

\section{Model and formalism}
\label{sec:model}
The model we consider in this work, i.e., the Hubbard atom, represents a basic building block of several fundamental Hamiltonians used for the description of many-electron physics. In particular, the Hubbard atom can be viewed as the limit of vanishing electronic hopping of the celebrated Hubbard model \cite{hubbard, esterling1970hubbard} or, equivalently, of vanishing hybridization in the not less famous Anderson impurity model \cite{anderson1961}. The atomic limit (AL) of the Hubbard model or Anderson Impurity model describes a single site with an instantaneous electrostatic repulsion $U$ in the presence of an external chemical potential $\mu$. Its explicit expression in second quantization formalism reads:
\begin{align}
\label{eq:H_HA}
    {\cal \hat{H}}= U \, \hat{n}_\uparrow \hat{n}_\downarrow -\mu \, (\hat{n}_\uparrow + \hat{n}_\downarrow),
\end{align}
 where the density operators of electrons with spin $\sigma=\uparrow, \downarrow$ can be expressed as $\hat{n}_\sigma=\hat{c}^\dagger_\sigma \hat{c}_\sigma$ in terms of the creation/annihilation operators $\hat{c}^\dagger_\sigma/\hat{c}_\sigma$ of an electron with spin $\sigma=\uparrow,\downarrow$ on the atomic site considered. Throughout this work, we will set $\mu=U/2$, which corresponds to fixing the average occupation of the system to $1$ electron (half-filling), i.e., to a regime where correlation effects are expected to be the strongest.
 
Due to the Pauli principle, the Hilbert space of the model consists of four states, corresponding to the state without electrons, i.e.~$|0\rangle$, to the two states with one electron in one of its two spin configurations  $|\!\uparrow\rangle$,$|\!\downarrow\rangle$, and to a double occupancy of electrons with opposite spins $|\!\uparrow \downarrow\rangle$. Evidently, this reduced Hilbert-space allows for an analytical solution and makes it possible to exactly compute several many-electron quantities typically used in the quantum many-body description of correlated electrons. 
At the same time, the presence of the interaction term $U$ allows to capture, albeit in a simplified fashion, nontrivial aspects of the many-electron physics in strong-coupling regimes, whose theoretical description is often quite hard.
In particular, among several studies, where the Hubbard atom has been used as a starting point for more complex analyses \cite{pairault2000, delre2021}, we should also recall that its solution \cite{pairault2000, thunstrom2018analytical, fus2022breakdown} turned out to be crucial to investigate the different manifestations of the breakdown of the self-consistent perturbation theory for the many-electron problem, such as the divergences of the two-particle irreducible vertex functions \cite{schafer2013divergent, janis2014MITtdivergence, schafer2016nonperturbative, chalupa2018, reitner2020attractive, springer2020, chalupa2021, adler2022, pelz2023highly, essl2023MA} as well as the misleading convergence of bold perturbative series \cite{kozik2015nonexistence, stan2015unphysical, tarantino2018nonperturbative, vucicevic2018practical, essl2023MA}, and to clarify their intrinsic interconnection \cite{gunnarsson2017breakdown}.

Here we focus on the generalized susceptibilities, which are directly related to the one- ($G^{(1)}$) and two-particle ($G^{(2)}$) Green's functions defined by
\begin{align}
    G^{(1)}_{\sigma}(\tau):=-\langle T_\tau \hat{c}_{\sigma}(\tau)\hat{c}^\dagger_{\sigma}(0)\rangle
    \label{eq:G1}
\end{align}
and
\begin{align}
     G^{(2)}_{\sigma_1\sigma_2\sigma_3\sigma_{4}}(\tau_1,\tau_2,\tau_3,0):=\langle T_\tau \hat{c}^\dagger_{\sigma_1}(\tau_1)\hat{c}_{\sigma_2}(\tau_2)\hat{c}^\dagger_{\sigma_{3}}(\tau_3) \hat{c}_{\sigma_{4}}(0)\rangle,
\label{eq:G2}
\end{align}
where the external brackets define the grand-canonical thermal averaging $\langle \cdots \rangle = \frac{1}{{\cal Z}} \mbox{Tr} \left[ e^{-\beta {\cal \hat{H}}} \cdots \right]$,  ${\cal Z}$ being the corresponding partition function and $\beta$ the inverse temperature of the system, $\hat{c}^{(\dagger)}_\sigma(\tau)=e^{\tau {\cal \hat{H}}}c^{(\dagger)}_\sigma e^{-\tau {\cal \hat{H}}}$, $T_\tau$ denotes  Wick's time-ordering operator in imaginary times $\tau$ and we have exploited the time-translational invariance of the problem considered, by setting the time argument of the last operator in the Green's function definitions equal to zero.

The specific quantities of interest for our work are Matsubara frequency dependent quantities defined by the Fourier transform as follows:
\begin{align}
G^{(1)}_\sigma(\nu) = \int_0^\beta d\tau \, e^{i \nu \tau } G^{(1)}_{\sigma}(\tau),
\end{align}
where the fermionic Matsubara frequencies $\nu$ are defined as $(2 n + 1) \frac{\pi}{\beta}$, $\forall n \in \mathbb{Z}$. We also introduce
the electronic self-energy defined as
\begin{align}
\Sigma_{\sigma}(\nu) = \left[G^{(1)}_{0,\sigma}(\nu)\right]^{-1} -  \left[G^{(1)}_{\sigma}(\nu)\right]^{-1},
\end{align}
where $G^{(1)}_{0,\sigma}(\nu) = \frac{1}{i\nu}$ is the Green's function of the corresponding noninteracting case ($U=\mu=0$). As no magnetic field is applied to our system ($h=0$), ${\cal \hat{H}}$ is perfectly $SU(2)$-symmetric in the spin-space, and all one-particle quantities, like $\Sigma_{\sigma}(\nu)$, are, thus, spin independent. For the half-filled Hubbard atom our convention for the chemical potential given below Eq.~\ref{eq:H_HA}  implies that the exact expression for the self-energy is $\Sigma(\nu)= \frac{U^2}{4i\nu}$ 

We are also interested in the susceptibilities defined from the two-particle correlator as:
\begin{align}
\label{ch_ph}
    \chi^{\nu\nu^\prime}_{\sigma\sigma^\prime}(\omega) =\int_0^\beta d\tau_1d\tau_2d\tau_3\;\chi_{\sigma\sigma^\prime}(\tau_1,\tau_2,\tau_3)\;e^{-i\nu\tau_1}e^{i(\nu+\omega)\tau_2}e^{-i(\nu^\prime+\omega)\tau_3}, 
\end{align}
where $\nu,\nu^\prime$ as well $\omega$ are respectively fermionic and bosonic Matsubara frequencies, the latter being defined as $ \frac{2 n \pi}{\beta}$, $\forall n \in \mathbb{Z}$, and 
\begin{align}
\label{genchidef}
     \chi_{\sigma\sigma^\prime}(\tau_1,\tau_2,\tau_3,0):=G^{(2)}_{\sigma\sigma\sigma^\prime\sigma^\prime}(\tau_1,\tau_2,\tau_3,0)-G^{(1)}_{\sigma}(\tau_1-\tau_2)G^{(1)}_{\sigma^\prime}(\tau_3).
\end{align}
The above definition of the generalized susceptibility in terms of this specific combination of one- and two-particle Green's function is particularly useful because after performing the summation on both fermionic Matsubara frequencies ($\nu, \nu'$), it directly yields the physical response of the system to an external perturbation in the framework of the linear response theory.

In particular, the following spin combinations define the \emph{generalized} charge ($c$)/spin ($s$) susceptibilities of our system:
\begin{align}
\chi^{\nu\nu^\prime}_{c/s}(\omega) = \chi^{\nu\nu^\prime}_{\uparrow\uparrow}(\omega) \pm \chi^{\nu\nu^\prime}_{\uparrow\downarrow}(\omega),
\label{eq:genchichsp}
\end{align}
where the $+ (-)$ sign should be used for the charge (spin) susceptibility, respectively. 
Hence, the corresponding \emph{physical} charge and spin susceptibilities can be directly computed as
\begin{align}
\chi_{c/s}(\omega) =\frac{1}{\beta^2} \sum_{\nu,\nu^\prime} \, \chi^{\nu\nu^\prime}_{c/s}(\omega).
\label{eq:physchi}
\end{align}
We recall that, in the (particularly relevant) static case ($\omega=0$), the two response functions yield respectively the isothermal compressibility ($\chi_c \propto \frac{\partial n}{\partial \mu}$) and the magnetic susceptibility ($\chi_s \propto \frac{\partial m}{\partial h}$) of the system considered. In the following, we will mostly focus on the static case, which amounts to set $\omega=0$ in Eq.~\eqref{ch_ph}, and will henceforth omit the bosonic frequency index. In this way the generalized susceptibilities of the Hubbard atom can be regarded as two-dimensional matrices in the fermionic Matsubara frequency space.

\section{Wavelet transform and methodological details}
\label{sec:wavelet}

The wavelet transform is a powerful tool for the analysis of frequency components in signals. Similar to the Fourier transform, the signal will be decomposed in a different basis set in frequency space where it is represented by coefficients obtained from the transform. Unlike the Fourier transform, the wavelet transform makes use of temporally enclosed, wave-like functions as its basis. These wavelet packets can be translated to gain information on the location in time of certain frequencies while scaling (stretching) wavelets allows a resolution in the frequency domain \cite{beylkin1991, daubechies1992ten, mallat1999wavelet}. Since its discovery, the advantages of the wavelet transform over other methods in frequency analysis have made it a popular tool in a wide range of areas such as image processing, data compression, or speech recognition \cite{beylkin1991,donoho1995,donoho1995vag,abramovich1998,mallat1999wavelet,candes2000,chirstopoulos2000,sifuzzaman2009application,akansu2010emerging}.

The transform computes coefficients by convolving sets of wavelet functions over a signal. Wavelets are wave-like functions $\psi \in \mathbf{L}^2(\mathbb{R})$ that share the following properties:
\begin{subequations}
\begin{align}
    &\int_{-\infty}^{\infty} \psi(t) dt = 0 \label{eq:wav_avgzero} ,\\
    \lVert \psi(t) \lVert^2 = &\int_{-\infty}^{\infty} \psi(t) \psi^*(t) dt = 1. \label{eq:wav_normone}
\end{align}
\end{subequations}
We can build a wavelet basis $\mathcal{D}$ by scaling and translating %a 
such a mother wavelet $\psi$:
\begin{equation}
\mathcal{D}=\left\{\psi_{u, s}(t)=\frac{1}{\sqrt{s}} \psi\left(\frac{t-u}{s}\right)\right\}_{u \in \mathbb{R}, s \in \mathbb{R}^{+}}
\end{equation}
where $u$ is used for translating the wavelet and $s$ determines the scale. The transformation of a function $f \in \mathbf{L}^2(\mathbb{R})$ is then defined as:
\begin{equation}\label{equ:psi_conv}
W f(u, s)=\int_{-\infty}^{+\infty} f(t) \frac{1}{\sqrt{s}} \psi^*\left(\frac{t-u}{s}\right) d t=f \star \bar{\psi}_s(u)
\end{equation}
with
\begin{equation}
\bar{\psi}_s(t)=\frac{1}{\sqrt{s}} \psi^*\left(\frac{-t}{s}\right)
\end{equation}
where $f \star \bar{\psi}_s(u)$ is the convolution
between $f$ and $\bar{\psi}_s(u)$. The Fourier transform, which henceforth will be denoted by a circumflex symbol over the quantity of interest, can be used to obtain $\bar{\psi}_s$ in the frequency domain
\begin{equation}\label{equ:ft_psi}
    \hat{\bar{\psi_s}}(\omega) = \sqrt{s}\hat{\psi}^*_s(\omega)
\end{equation}
From Eq.~\ref{eq:wav_normone} one can see that $\hat{\psi}(0) = \int_{-\infty}^{+\infty} \psi(t) dt = 0$. Hence the wavelet function $\psi$ can be interpreted as a high pass filter \cite{mallat1999wavelet}.

The inverse wavelet transform of a function $f(t)$ can be defined as 
\begin{equation}
    f(t)= \frac{1}{C_{\psi}}\int_0^{+\infty} \int_{-\infty}^{+\infty} W f(u, s) \frac{1}{\sqrt{s}} \psi\left(\frac{t-u}{s}\right) d u \frac{ds}{s^2}
\end{equation}
where $C_{\psi}$ is given by
\begin{equation}\label{equ:admiss}
    C_{\psi} = \int_0^{+\infty} \frac{|\hat{\psi(\omega)}|^2}{\omega} < \infty
\end{equation}
The above equation is also known as weak admissability condition \cite{mallat1999wavelet}. 

We can introduce a \emph{scaling function} $\phi$, also known as father wavelet, which captures the low-frequency components of the signal representing slower changes in the data. In other words this corresponds to an accumulation of wavelet functions for scales $s$ larger than $1$. The scaling function is essential for the reconstruction of the original signal. It provides the basis for the low-frequency components or approximation at each level of resolution, while the wavelet functions capture the detail components. The modulus of the Fourier transform of $\phi$ can be computed using the Fourier transform of the mother wavelet as:
\begin{equation}
|\hat{\phi}(\omega)|^2=\int_1^{+\infty}|\hat{\psi}(s \omega)|^2 \frac{d s}{s}=\int_\omega^{+\infty} \frac{|\hat{\psi}(\xi)|^2}{\xi} d \xi
\end{equation}
from which the scaling function can then be derived via an anti Fourier transform \cite{mallat1999wavelet}. Using Eq.~\eqref{equ:admiss} and Eq.~\eqref{eq:wav_normone} we see that $\lim_{\omega\to0} |\hat{\phi}(\omega)|^2 =  C_{\psi}$, showing that the scaling function can be interpreted as a low pass filter. Furthermore, defining the scaling function as 
\begin{subequations}
\begin{align}
    \phi_s(t) = \frac{1}{\sqrt{s}}\phi_s(\frac{t}{s})\\
    \bar{\phi}_s (t) = \phi_s^*(-t)
\end{align}    
\end{subequations}
allows us, in analogy to Eq.~\eqref{equ:psi_conv}, obtain the low frequency approximation of $f$ at scale $s$ via a convolution with the scaling function $f \star \bar{\phi}^*_s(u)$ \cite{mallat1999wavelet}.

\begin{figure}[!htb]
    \centering
    \begin{subfigure}{0.48\textwidth}
        \centering
        \includegraphics[width=\linewidth]{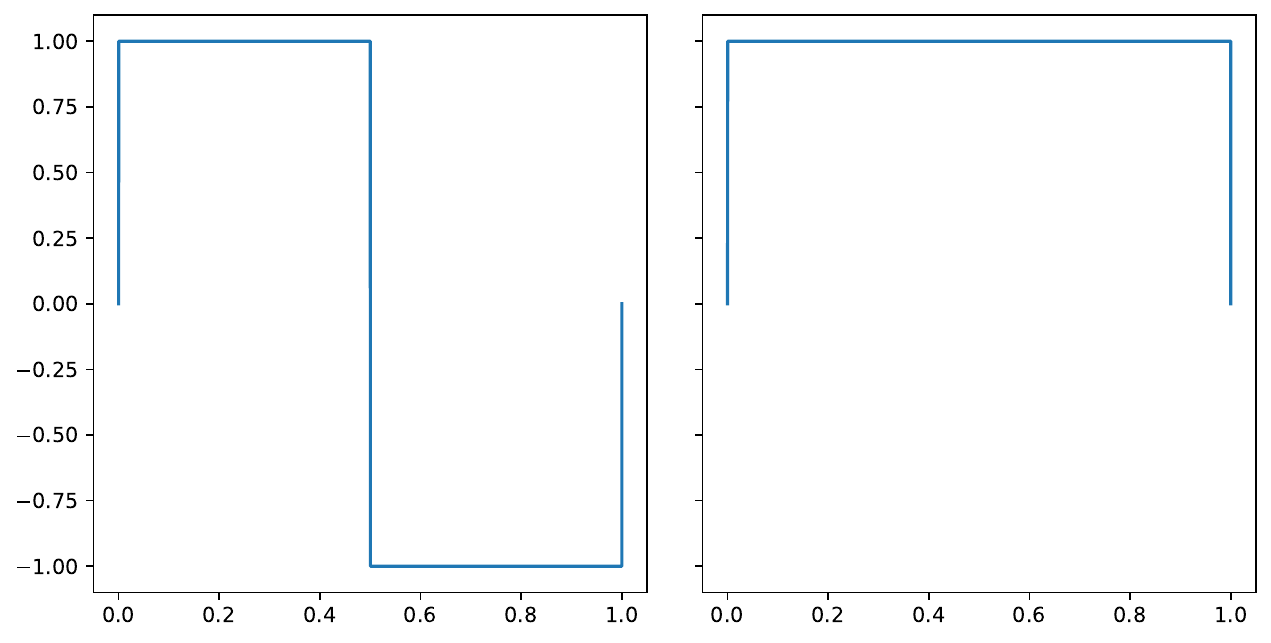}
        \caption{Haar}
    \end{subfigure}
    \hfill
    \begin{subfigure}{0.48\textwidth}
        \centering
        \includegraphics[width=\linewidth]{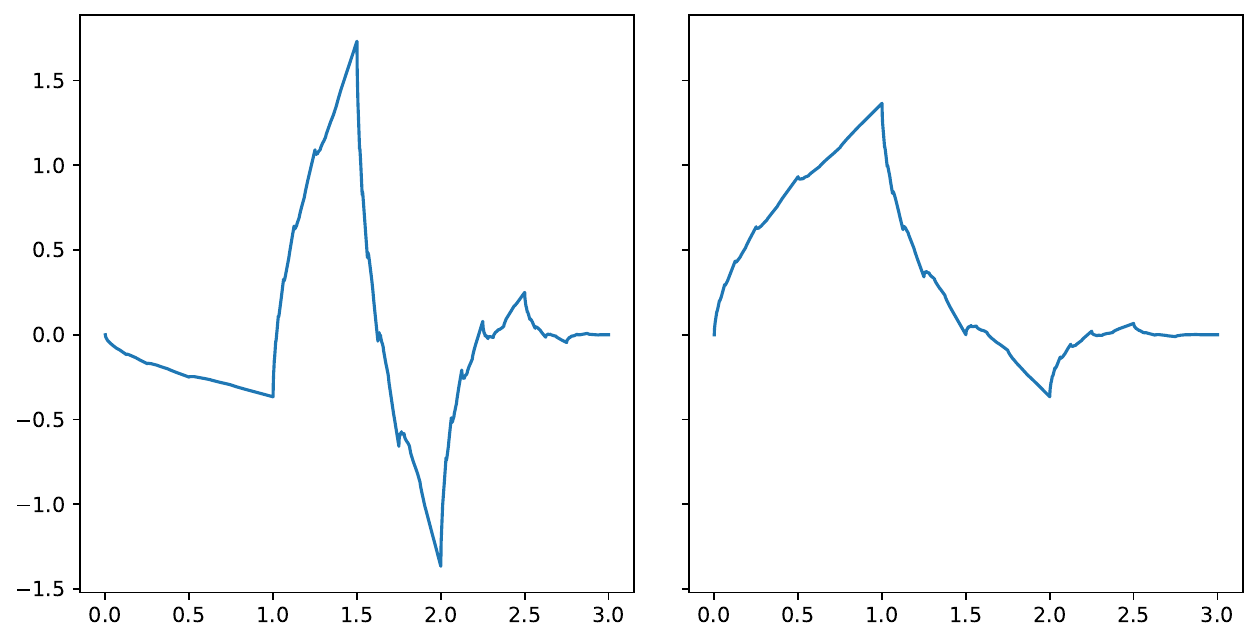}
        \caption{Daubechies 2}
    \end{subfigure}
    \par\bigskip
    \begin{subfigure}{0.48\textwidth}
        \centering
        \includegraphics[width=\linewidth]{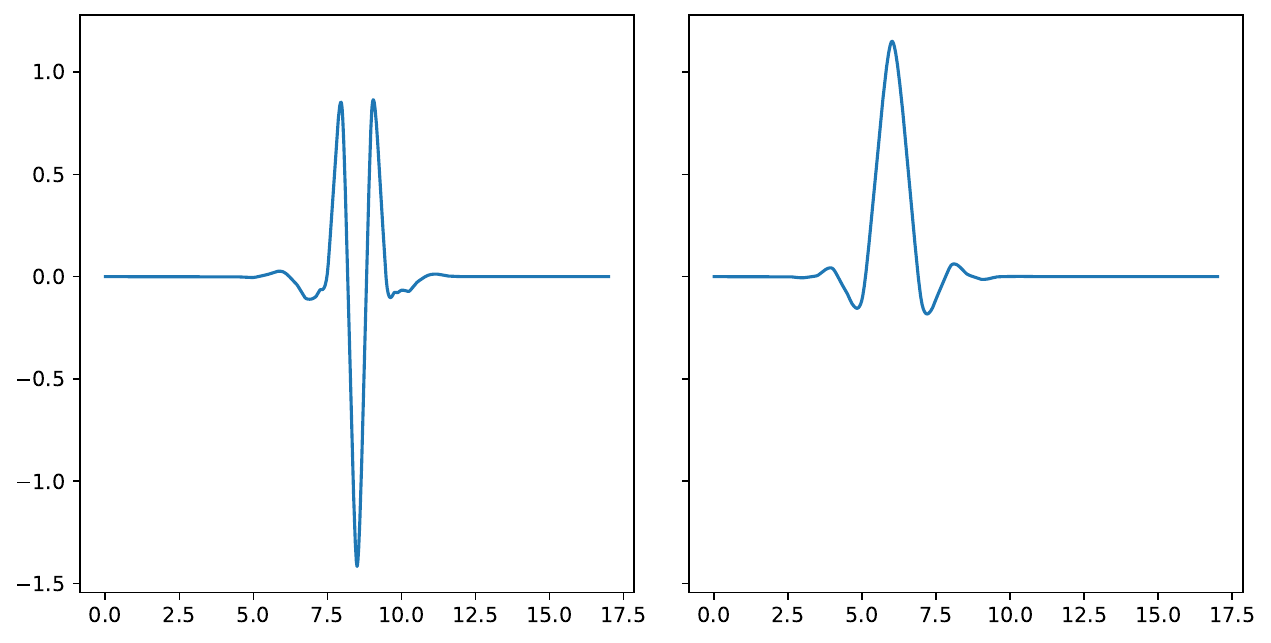}
        \caption{Coiflet 3}
    \end{subfigure}
    \hfill
    \begin{subfigure}{0.48\textwidth}
        \centering
        \includegraphics[width=\linewidth]{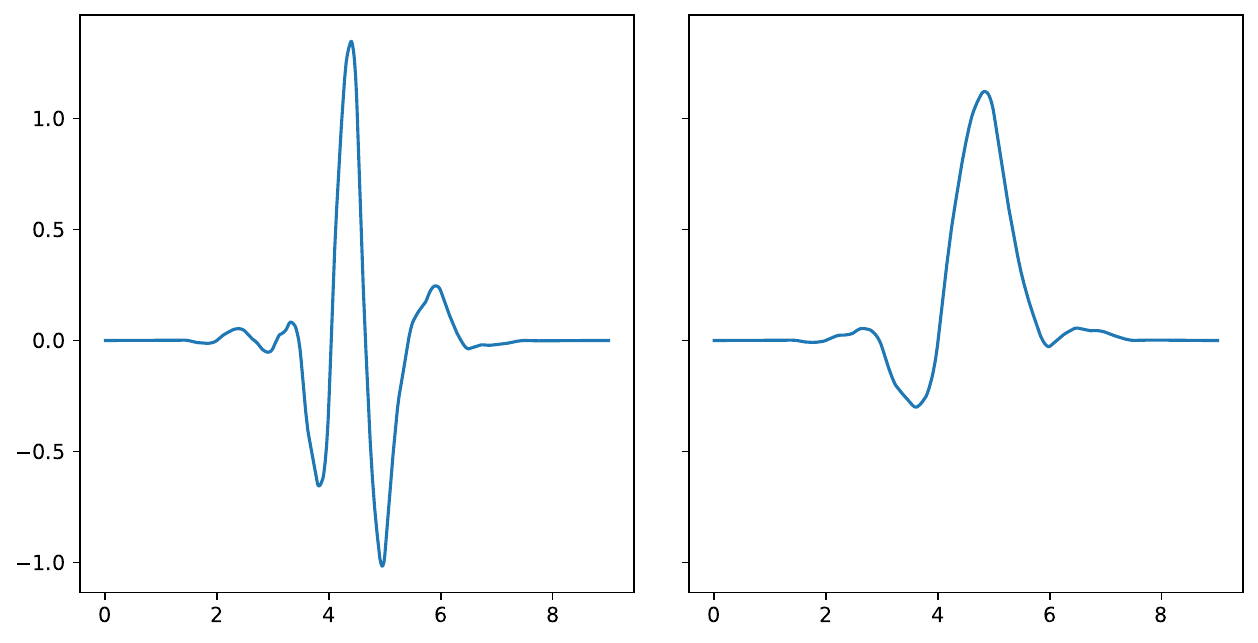}
        \caption{Symlet 5}
    \end{subfigure}
    \caption{Wavelet functions $\psi$ (left) and scaling functions $\phi$ (right) of different wavelet bases.}
    \label{fig:wavelets_shape}
\end{figure}

Fig.~\ref{fig:wavelets_shape} examples for prominent wavelet functions $\psi$ and the corresponding scaling function $\phi$ of different wavelets are shown.

In order to practically execute computations the discrete wavelet transform (DWT) \cite{beylkin1991, daubechies1992ten, mallat1999wavelet}, is used. Using the DWT, a discrete signal $f$ can be rewritten as follows \cite{chun2010tutorial}:
\begin{equation}
f[n]  =\frac{1}{\sqrt{N}} \sum_k W_\phi\left[j_0, k\right] \phi_{j_0, k}[n] +\frac{1}{\sqrt{N}} \sum_{j=j_0}^{\infty} \sum_k W_\psi[j, k] \psi_{j, k}[n]
\end{equation}
Observe, that we have two sets of basis functions $\{\phi_{j_0,k} [n] \}_{k\in\mathbb{Z}}$ and $\{\psi_{j,k} [n] \}_{(j,k)\in\mathbb{Z}^2}$ where $j$ and $k$ denote the indices associated to scaling and translating the wavelets. The functions $\phi[n]$, $\psi[n]$ and $f[n]$ are defined on a discrete interval $[0, N-1]$. As before, the function $\phi$ corresponds to the scaling function, while $\psi$ refers to the wavelet function. Since these two functions are orthogonal to each other, we can compute the coefficients $W_i$ as discrete convolutions of the functions $\psi$ and $\phi$ over the discretized signal $f$ \cite{chun2010tutorial}:
\begin{subequations}
\begin{align}
W_\phi\left[j_0, k\right] & =\frac{1}{\sqrt{N}} \sum_n f[n] \phi_{j_0, k}[n],  \\
W_\psi[j, k] & =\frac{1}{\sqrt{N}} \sum_n f[n] \psi_{j, k}[n] .
\end{align}
\end{subequations}

As mentioned in Sec.~\ref{sec:model}, since we are mainly interested in the static ($\omega=0$) case, we can represent the two-particle quantities as matrices depending on the \emph{two} fermionic frequencies $\nu, \nu'$. Consistently, a two-dimensional DWT can be defined by applying the one-dimensional DWT first on the rows of our signal and subsequently on the columns \cite{beylkin1991, mallat1999wavelet, daubechies1992ten}.  In this way we replace the original one-dimensional scaling and wavelet functions by their two-dimensional counterparts $\{\phi, \psi^h$, $\psi^v$, $\psi^d\}$, where the former is now the two-dimensional scaling function responsible for capturing the low frequency components of the data and the latter three, which denote the two-dimensional wavelet functions, are responsible for the details in the horizontal, vertical and diagonal directions of the signal, respectively. Hence, the corresponding two-dimensional signal $f$, indexed by two indices $m$ and $n$, can now be represented by \cite{chun2010tutorial}
\begin{equation}\label{equ:2d_dwt}
\begin{aligned}
f[m, n]&=\frac{1}{\sqrt{M N}} \sum_k \sum_r W_\phi\left[j_0, k, r\right] \phi_{j_0, k, r}[m, n] + \frac{1}{\sqrt{M N}} \sum_{i \in\{h, v, d\}} \sum_{j=j_0}^{\infty} \sum_k \sum_r W_\psi^i[j, k, r] \psi_{j, k, r}^i[m, n] 
\end{aligned}
\end{equation}
where now an additional index $r$ has been added for translations along the second dimension. As before, the coefficients $W$ can be obtained by convolving the signal $f$ with the scaling and wavelet functions \cite{chun2010tutorial}:
\begin{subequations}\label{equ:2d_coeff}
\begin{align}
W_\phi\left[j_0, k, r\right]=&\frac{1}{\sqrt{M N}} \sum_{m=0}^{M-1} \sum_{n=0}^{N-1} f[m, n] \phi_{j_0, k, r}[m, n] ,\\
W_\psi^i[j, k, r]=&\frac{1}{\sqrt{M N}} \sum_{m=0}^{M-1} \sum_{n=0}^{N-1} f[m, n] \psi_{j, k, r}^i[m, n], 
\end{align}
\end{subequations}
where $i \in\{h, v, d\}$. If we consider the specific case of our interest, i.e.. the two-dimensional DWT on the generalized susceptibilities, we can obtain the corresponding explicit expressions  simply by  making the substitutions $f \to \chi$ and $m, n \to \nu, \nu'$ in  Eqs.~\eqref{equ:2d_dwt} and \eqref{equ:2d_coeff}. Evidently, the DWT outlined above can be easily generalized to arbitrary dimensions (e.g. to 3D, if one needs to compress bosonic-frequency dependent susceptibilities) by simply applying the transform consecutively along each dimension of the signal \cite{mallat1999wavelet}.

The wavelet transform can be applied to a signal multiple times recursively. The depth of this recursion is denoted by the \emph{level} of the decomposition. Each computation of the decomposition at level $l+1$ takes the approximation coefficients of the current level $l$ and applies the decomposition to this approximation \cite{ mallat1989theory, beylkin1991, daubechies1992ten, mallat1999wavelet, chun2010tutorial}.
\begin{figure}
  \centering

  \begin{subfigure}{0.48\textwidth}
    \centering
    \includegraphics[width=\linewidth]{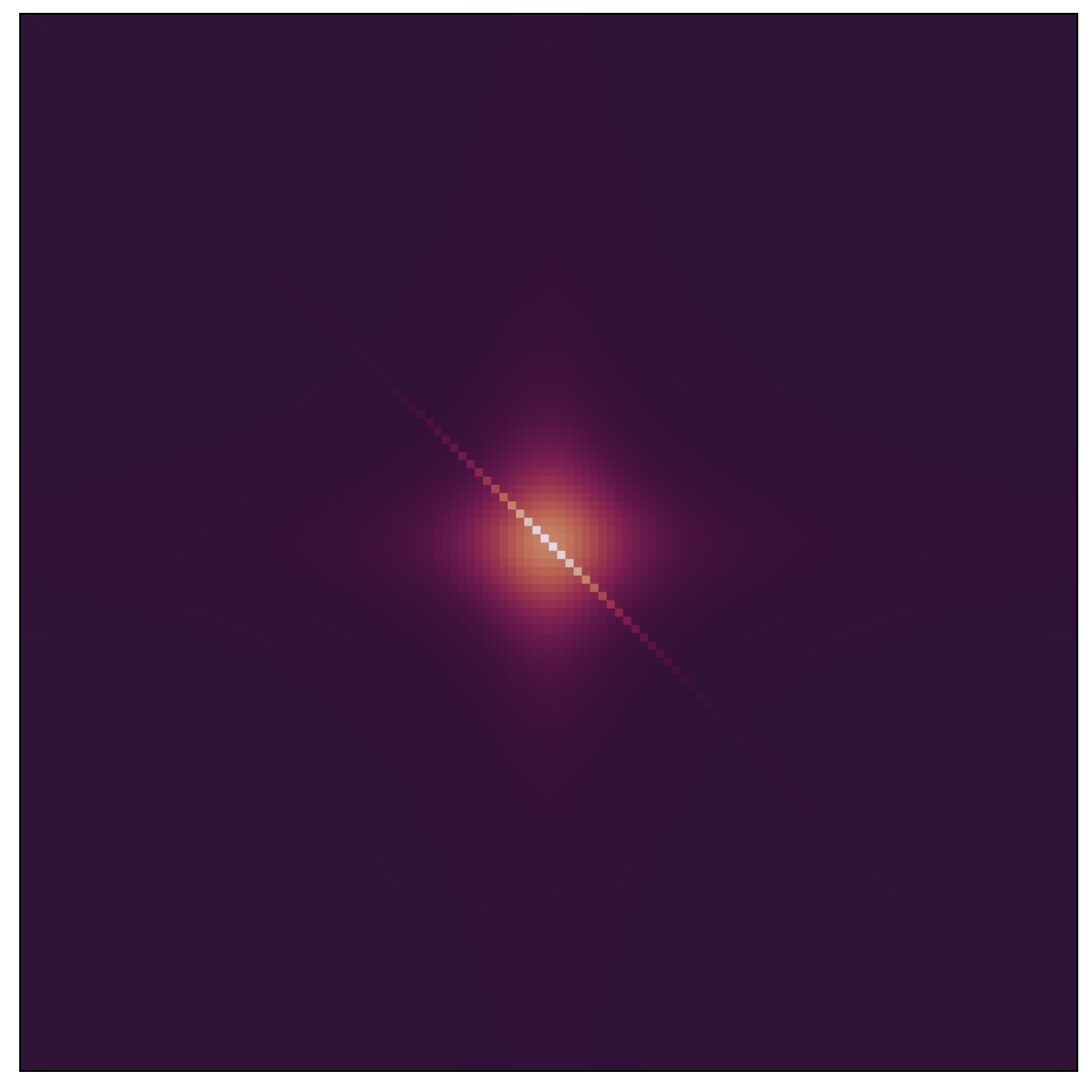}
    \caption{Original Vertex}
    \label{fig:wt_original}
  \end{subfigure}
  \hfill
  \begin{subfigure}{0.48\textwidth}
    \centering
    \includegraphics[width=\linewidth]{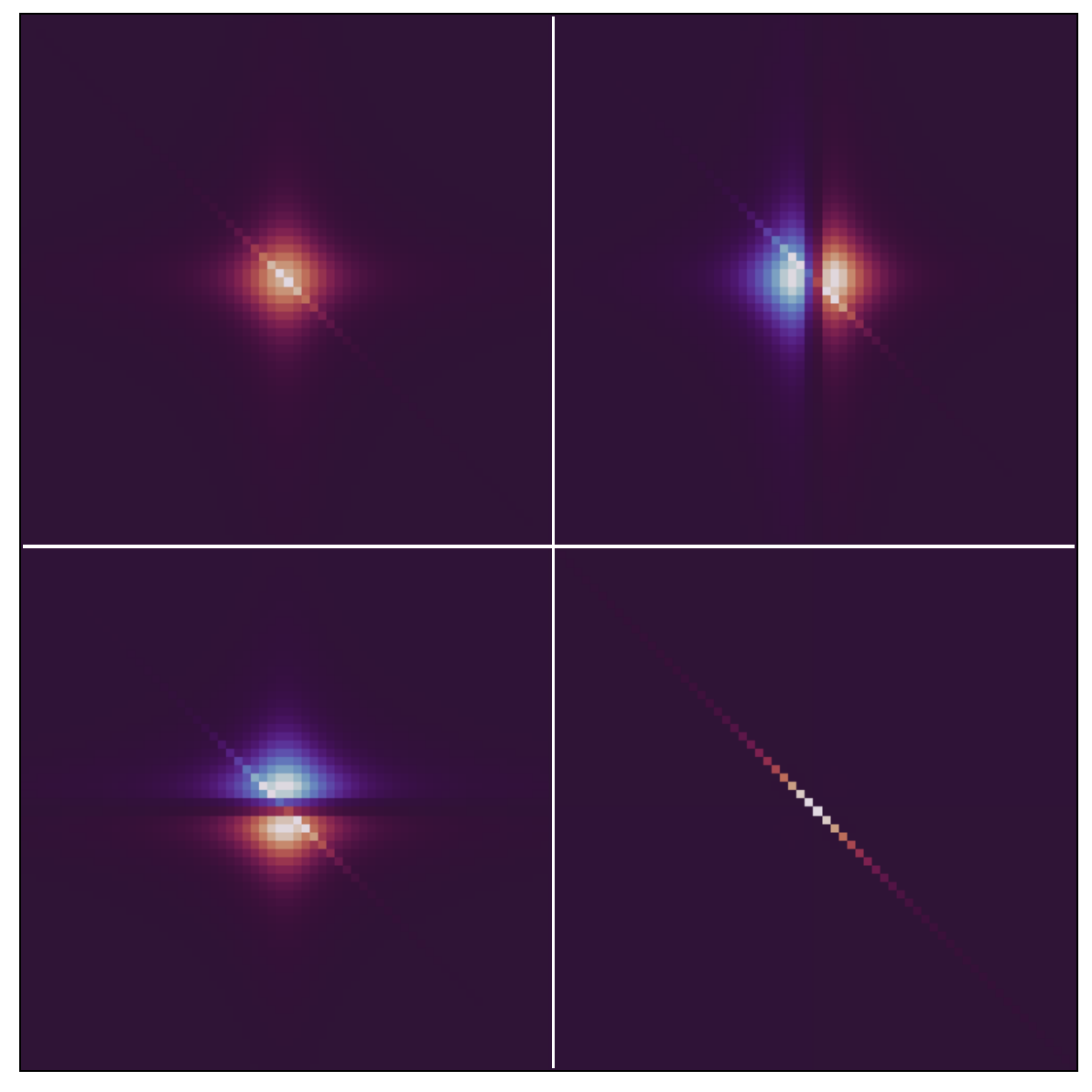}
    \caption{Level 1 Decomposition}
    \label{fig:wt_lev1}
  \end{subfigure}

    \par\bigskip

  \begin{subfigure}{0.48\textwidth}
    \centering
    \includegraphics[width=\linewidth]{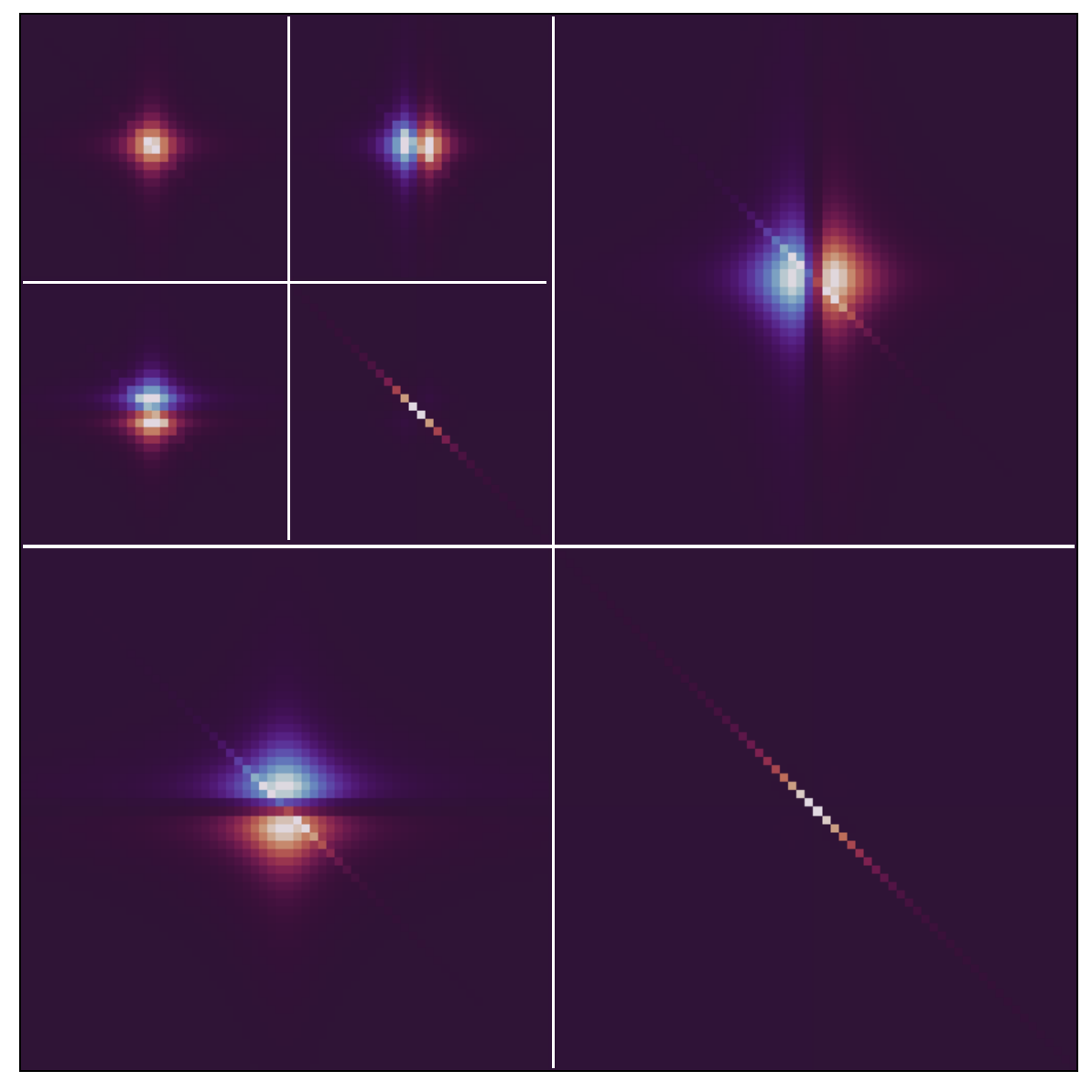}
    \caption{Level 2 Decomposition}
    \label{fig:wt_lev2}
  \end{subfigure}
  \hfill
  \begin{subfigure}{0.48\textwidth}
    \centering
    \includegraphics[width=\linewidth]{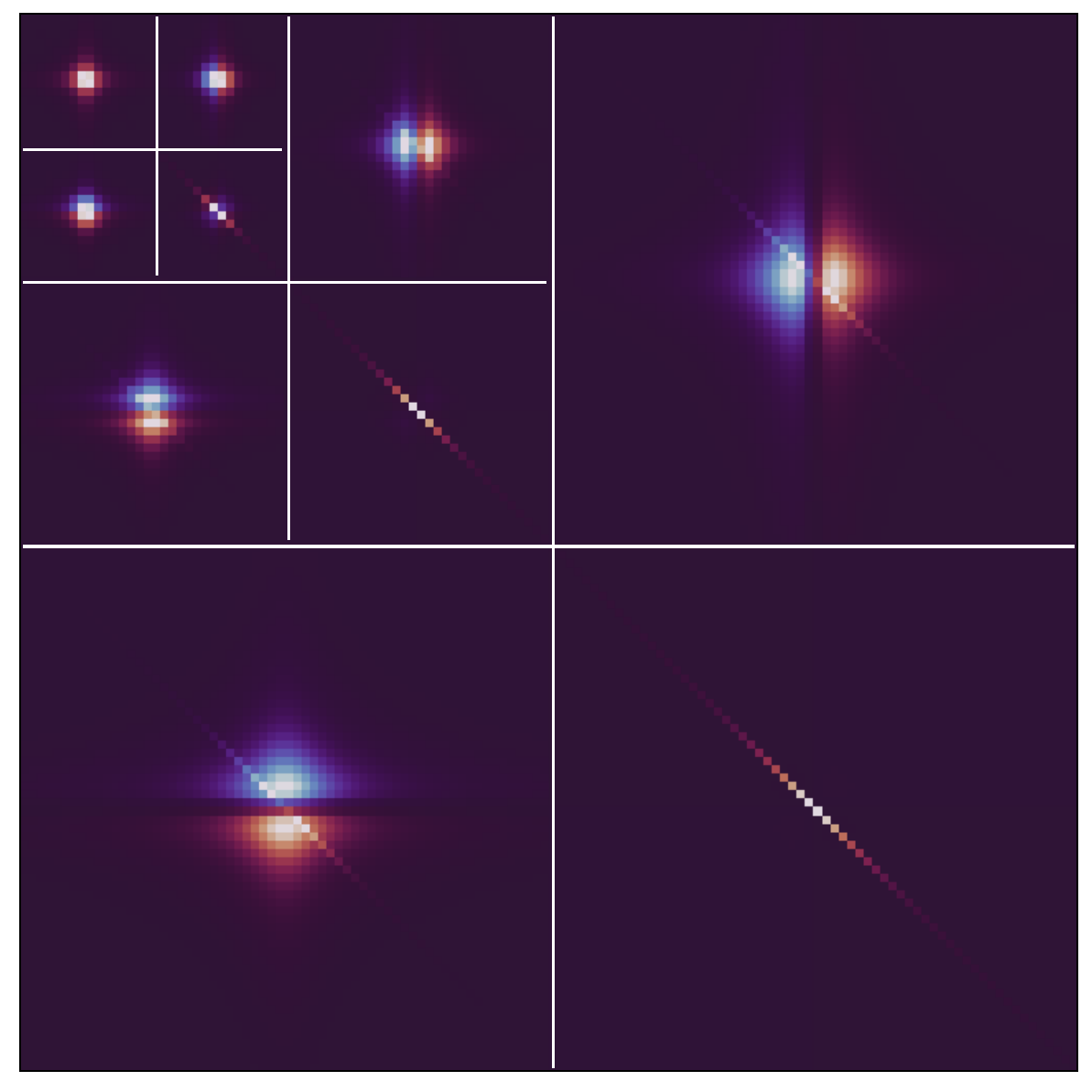}
    \caption{Level 3 Decomposition}
    \label{fig:wt_lev3}
  \end{subfigure}

  \caption{Decomposition of the generalized spin susceptibility vertex at $\beta = 100$, using the Haar wavelet (the coefficients of the wavelet decomposition are normalized for better visibility).}
  \label{fig:wt_demo}
\end{figure}
In Fig.~\ref{fig:wt_demo} the wavelet transform applied to an image for different levels is displayed. Fig.~\ref{fig:wt_lev1} shows a wavelet decomposition of level 1. The top left quarter is the approximation, while the top right, bottom left and bottom right quarters correspond to the horizontal, vertical and diagonal detail coefficients, respectively. Fig.~\ref{fig:wt_lev2} and Fig.~\ref{fig:wt_lev3} illustrate the recursion where the approximations that can be seen in the top left corners of Fig.~\ref{fig:wt_lev1} and Fig.~\ref{fig:wt_lev2} are replaced by, again, another set of approximation and detail coefficients.

In the following, we introduce the key methodological details required for obtaining the results presented in the next section. First, we describe how the wavelet transform is used to compress the data we are working with. Afterwards, we will highlight which metrics can be exploited to quantify the compression.

\subsection{Compression using wavelet transform}

We use the discrete wavelet transform as a means of compression as implemented in the open source package \emph{PyWavelets} \cite{Lee2019}. In contrast to other popular methods such as the Principal Component Analysis (PCA) \cite{ringner2008principal} or the Singular Value Decomposition (SVD) \cite{stewart1993early} we rely on a wavelet basis that is agnostic to the input data.

\begin{figure}[htbp]
    \centering
    \includegraphics[width=0.8\linewidth]{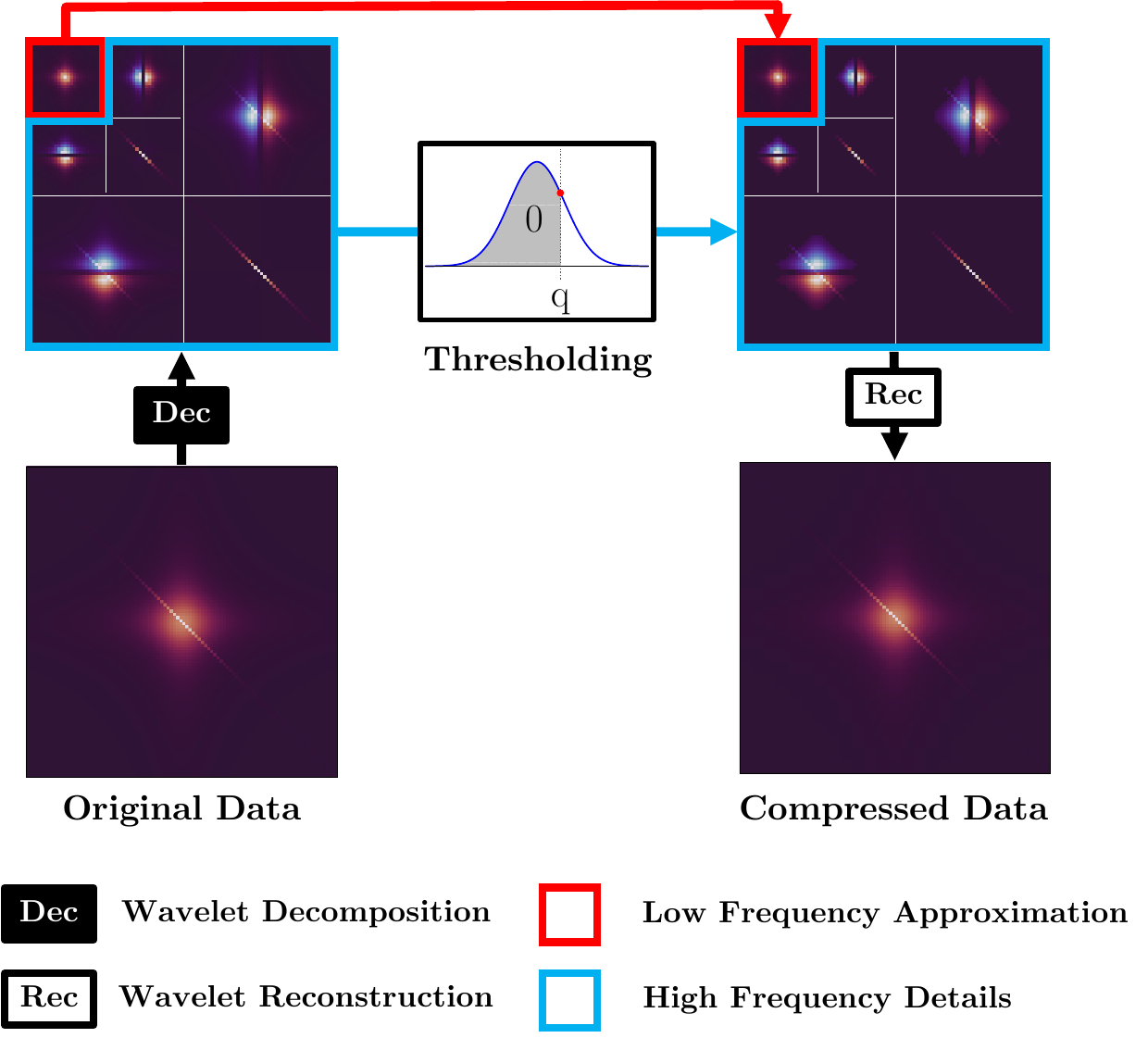}
    \caption{The proposed compression algorithm using the wavelet transform. After the wavelet decomposition of the original data, the absolute high-frequency detail coefficients are thresholded by a given $q$-quantile: all coefficients whose magnitude is smaller than the quantile will be set to zero. The wavelet reconstruction yields then the compressed data. Figure taken from \cite{dräger2023bulletin}.}
    \label{fig:compression_algo}
\end{figure}

The compression method we chose is rather simple and brute force. For a decomposition of level $l$, we leave the highest level approximation as it is. All the detail coefficients will be truncated based on their position in the distribution of detail coefficients. More specifically, we compute the $q$-quantile $d_q$ over the distribution of absolute values of the detail coefficients and set all detail coefficients $d$ where $\abs{d} \le d_q$ to $0$. The reconstruction of the sparsified coefficients then yields the compressed reconstruction of the original data. The proposed procedure is illustrated in Fig. \ref{fig:compression_algo}.

\subsection{Quantification of compression}\label{sec:quantification}

In order to compare the effect of the compression on the physical quantities as well as to assess the compression strength with varying degrees of decomposition levels and threshold quantiles, we first need to introduce a reliable measure. For our purposes, we choose two different metrics for the quantification of the wavelet-based compression. The first method we use, permits us to measure the efficiency of our compression algorithm in terms of memory requirements. The second metric, which is based on established methods in image processing, is used to quantify to which degree the compressed image deviates from the original one.

\subsubsection{Memory usage}
The original vertices can be represented as dense matrices. Almost each element of the matrix is nonzero. In memory, this dense matrix is stored as a contiguous block of numeric values, each corresponding to one entry in the matrix. If we have an $n \times m$ matrix, we occupy $nm$ blocks of the size of the datatype that stores the value of each entry. In our case, the values are represented in double-precision floating point format occupying 8 bytes.

When compressing the matrix using the wavelet transform, we expect the majority of entries to become zero transforming the dense matrix into a rather sparse one, depending on the strength of the compression. Since most of the elements then share the same value of 0, the matrix can be stored in a more memory efficient format. One of the most simple formats to achieve this efficiency is the COO format as described in \cite{shahnaz2005review}, where instead of storing a numeric value for each entry, we store tuples \emph{only} for those values that are nonzero. To characterize a nonzero entry and reconstruct a dense matrix representation, we require three values for each nonzero entry: the row and the column in the original representation and the value itself.

To quantify the level of compression, we simply compute a compression factor from the space taken up in memory. The size in memory of the original matrix in dense matrix representation $m_o$ and memory size of the reconstructed matrix in sparse matrix format $m_r$ form a compression ratio $\frac{m_r}{m_o}$. In the remainder of this work, however, we are interested in the percentage of compression, that the algorithm achieves. This \emph{relative compression ratio} can be expressed via
\begin{equation}\label{equ:CR}
    c = 1 - \frac{m_r}{m_o}.
\end{equation}
Note, that the representation of rather dense matrices in sparse matrix format might be larger than in the dense matrix format due to overhead in the data structure.

\subsubsection{Structural similarity index measure (SSIM)}

The sheer memory size of the vertices is essentially a measure of sparsity in the matrix. To get better insight into the compression, we adopt the Structural Similarity Index Measure (SSIM) into our analysis. The SSIM, which was first introduced by Wang \emph{et al.} in Ref.~\cite{wang2004image}, is a measure for the similarity of two images and is widely used in image processing. \cite{dosselmann2011comprehensive, nilsson2020understanding}

The SSIM is defined as
\begin{equation}
\operatorname{SSIM}(x, y)=(l(x, y))^\alpha(c(x, y))^\beta(s(x, y))^\gamma,
\end{equation}
where $x$ and $y$ refer to the coordinates in the input images $\mathbf{A}$ and $\mathbf{B}$. The quantities $l$, $c$ and $s$ denote the so-called \emph{luminance}, \emph{contrast} and \emph{structure}, respectively. They are computed using an $11 \times 11$ pixel Gaussian filter around $x$ and $y$ with a standard deviation of $\sigma=1.5$ \cite{wang2004image}.

The SSIM essentially splits the task of comparing the similarity of two images into three subtasks: (i) a comparison of the luminance, (ii) a comparison of the contrast, and (iii) a comparison of the structure of the images before combining their results to obtain the final index \cite{wang2004image}. The luminance is defined as
\begin{equation}
l(x, y)=\frac{2 \mu_{\mathbf{A}} \mu_{\mathbf{B}}+C_1}{\mu_{\mathbf{A}}^2+\mu_{\mathbf{B}}^2+C_1},
\end{equation}
where $\mu_{\mathbf{A}}$ and $\mu_{\mathbf{B}}$ are the mean values computed on the input images and are functions of $x$ and $y$. Similarly, we define $\sigma_{\mathbf{A}}$ and $\sigma_{\mathbf{B}}$ as the variances on the input images to compute the contrast $c$ as
\begin{equation}
c(x, y)=\frac{2 \sigma_{\mathbf{A}} \sigma_{\mathbf{B}}+C_2}{\sigma_{\mathbf{A}}^2+\sigma_{\mathbf{B}}^2+C_2}.
\end{equation}
Finally, the structure $s$ is defined in a very similar way as
\begin{equation}
s(x, y)=\frac{\sigma_{\mathbf{A B}}+C_3}{\sigma_{\mathbf{A}} \sigma_{\mathbf{B}}+C_3},
\end{equation}
where $\sigma_{\mathbf{A B}}$ now represents the covariance. For simplicity we set $\alpha = \beta = \gamma = 1$ and $C_3 = \frac{C_2}{2}$ to obtain the SSIM \cite{wang2004image, nilsson2020understanding, brunet2011mathematical}:
\begin{equation}
\operatorname{SSIM}(x, y)=\frac{\left(2 \mu_{\mathbf{A}} \mu_{\mathbf{B}}+C_1\right)\left(2 \sigma_{\mathbf{A B}}+C_2\right)}{\left(\mu_{\mathbf{A}}^2+\mu_{\mathbf{B}}^2+C_1\right)\left(\sigma_{\mathbf{A}}^2+\sigma_{\mathbf{B}}^2+C_2\right)}.
\end{equation}
Here, $L$ corresponds to the dynamic range values for each pixel. For an 8-bit image we have $L=255$. $C_1$ and $C_2$ are then defined as $C_1=(K_1 L)^2$ and $C_2 = (K_2 L)^2$ with $K_i \ll 1$. In Ref.~\cite{wang2004image} $K_1$ and $K_2$ were set to $K_1=0.01$ and $K_2=0.03$.

%--------------------------------------------------------------------------------------------------------
%----------------------------------------- RESULTS ------------------------------------------------------
%--------------------------------------------------------------------------------------------------------

\section{Results}\label{sec:results}
For the quantification of our proposed wavelet-based compression algorithm we will now present and discuss a series of numerical experiments on the generalized, static susceptibilities of the Hubbard atom at half filling (see Sec.~\ref{sec:model}, Eqs.~\eqref{ch_ph}-\eqref{eq:physchi}). The experiments are chosen with the purpose to highlight characteristic physical properties of the model and how these are affected by the discrete wavelet transform (DWT). As a prototypical test bed for our compression procedure, we use \emph{Haar} wavelets as the basis of our choice throughout this work. In the Appendix \ref{sec:app_gridsize} we will report selected results showcasing other wavelet families.

\subsection{Compression of generalized charge and spin susceptibilities}

\subsubsection{Quantification of the compression}\label{sec:compr}

\begin{figure}
    \centering
    \includegraphics[width=\textwidth]{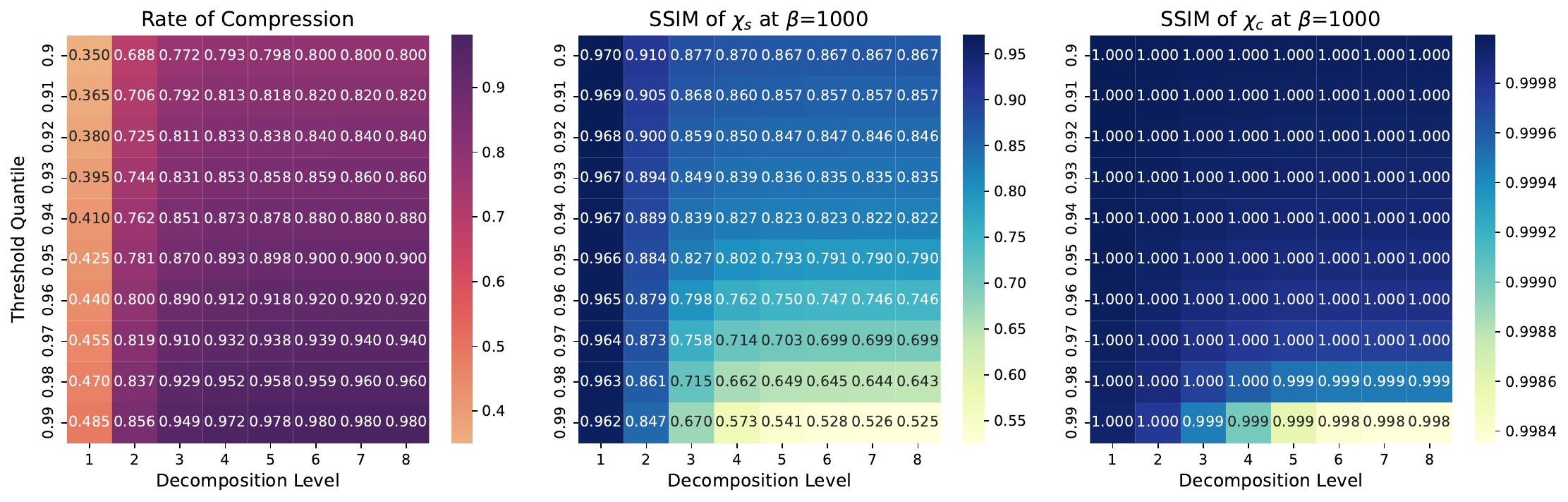}
    \caption{Left panel: Relative compression ratio $1-m_r/m_o$ for the generalized susceptibilities $\chi^{\nu, \nu'}$ as a function of decomposition level and threshold quantile. Note, that due to our chosen compression procedure, the compression rate is independent of temperature or which channel, i.e. spin or charge, is chosen. Central panel: SSIM as a function of the decomposition level and threshold quantile for the generalized spin susceptibility at $U=1$ and an inverse temperature of $\beta=1000$. Right panel: SSIM for the generalized charge susceptibility at $U=1$ and $\beta=1000$. For the calculations here a frequency grid size of $512\times512$ was chosen. For lower values of $\beta$ the SSIM scores for the spin and charge channel show close to no deviation from 1 independently of threshold quantile or decomposition level.}
    \label{fig:cr_ssim}
\end{figure}

\begin{figure}[h!]
    \centering
    \includegraphics[width=\textwidth]{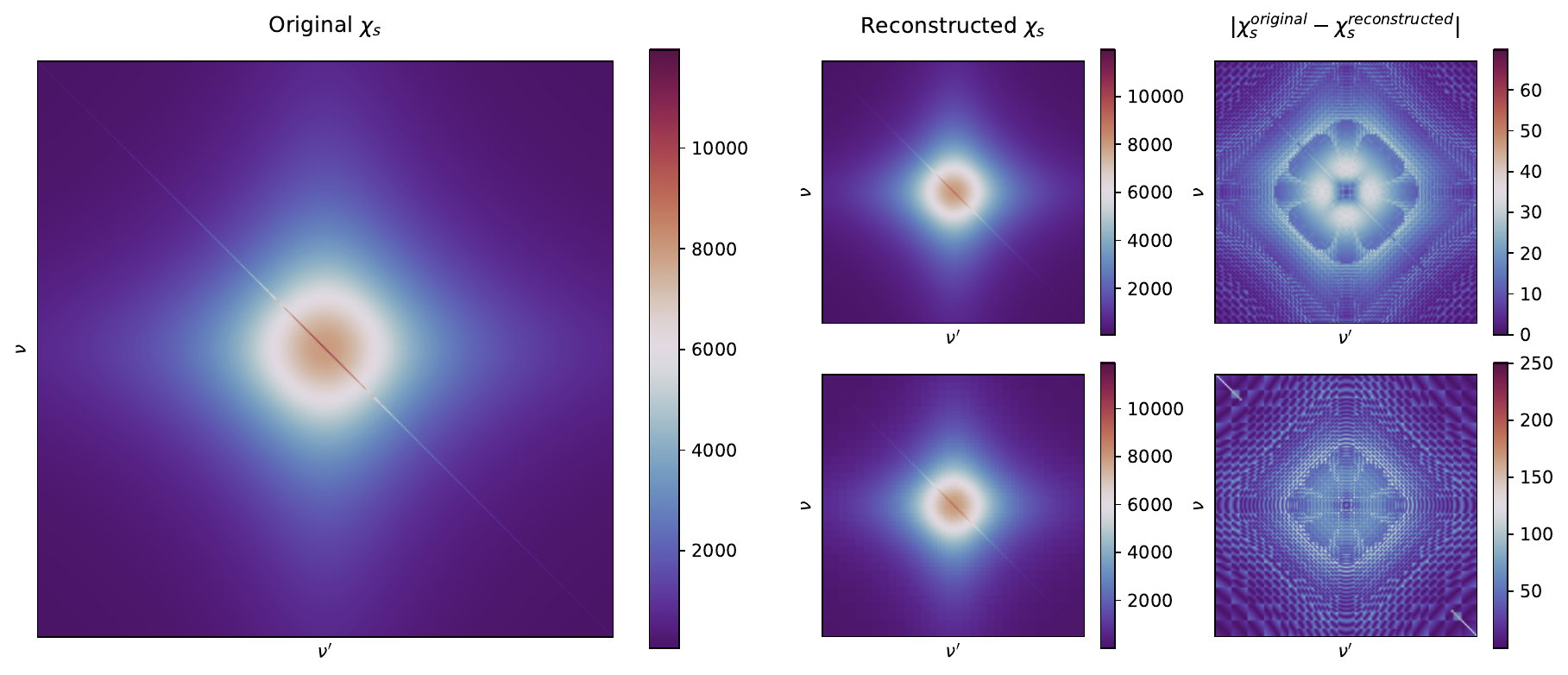}
    \caption{Illustrative example showcasing the wavelet compression of the generalized spin susceptibility $\chi_s^{\nu \nu'}$ for a Hubbard atom at half-filling, for $\beta=1000$ and $U=1$. Left panel: Original susceptibility. Central panel: Reconstructed susceptibility after a compression at a maximal possible decomposition Level $l=8$ and threshold quantile of $q=0.95$ (upper row) and $q=0.99$ (lower row). Right panel: Absolute error between the original and reconstructed susceptibilities for $q=0.95$ and $q=0.99$.}
    \label{fig:compr_compare}
\end{figure}

We begin our discussion with the analysis of the compression strength of the DWT applied to the generalized spin and charge susceptibilities of the Hubbard atom. To this end, we will compress the atomic limit vertices using various decomposition levels and threshold quantiles and afterwards compare the reconstructed quantities to the original ones. Using the two metrics discussed in Sec.~\ref{sec:quantification}, namely the \emph{relative compression ratio} and the SSIM, we analyse the performance of the DWT in terms of sheer memory requirements as well as reconstruction quality. The two-particle quantities of the atomic limit were evaluated for $\beta=1000$ and $U=1$ where a frequency grid size of $512\times512$ was chosen. The results are presented in Fig.~\ref{fig:cr_ssim}. The left panel shows the relative compression ratio, or $1-m_r/m_o$, as a function of decomposition level and threshold quantile. Evidently, with increasing threshold quantile and decomposition level the memory compression increases. We note, that since we truncate a fixed percentage of detail coefficients during our compression procedure, the rate of compression is independent of temperature as well as of which channel, i.e. spin or charge, we consider. The central and right panels show the SSIM scores for the spin and charge channel respectively again as a function of decomposition level and threshold quantile. We observe a clear-cut difference in performance between the two quantities. In particular, whereas the charge susceptibility displays a nearly perfect score across the board, the reconstruction of the spin susceptibility visibly deteriorates for higher decomposition levels and threshold quantiles. The above calculations were also conducted for lower $\beta$ values, namely for $\beta = \{10, 100\}$. In these cases, however, both for the spin and charge susceptibilities the wavelet compression achieved an extremely high SSIM score, which was almost indistinguishable from $1$. In order to provide a potentially more qualitative insight into the results presented in Fig.~\ref{fig:cr_ssim}, we show explicit examples of the wavelet compression for the specific case of the spin channel in Fig.~\ref{fig:compr_compare}. In the left panel we plot the spin susceptibility at $U=1, \beta=1000$ which are compared to two reconstructed susceptibilities at decomposition levels $l=8$ and threshold quantiles $q=0.95$ and $q=0.99$ which are respectively shown in the upper and lower row of the central panel. In the graphs of the right panel the absolute error between the original and reconstructed quantities is presented. 
The difference of the scale for the two cases indicates that an increasing threshold quantile or rather a stronger compression will result in a larger reconstruction error, consistent with the SSIM scores displayed in Fig.~\ref{fig:cr_ssim}.

We also notice here that differently from other schemes, our DWT method does not suffer from the shift of sharp features in the $\nu - \nu'$ plane (see diagonal structure in Fig.~\ref{fig:wt_original}) occurring for a finite bosonic frequency $\omega$.

\subsubsection{Preservation of symmetries}\label{sec:sym}

\begin{figure}
    \centering
    \includegraphics[width=0.8\textwidth]{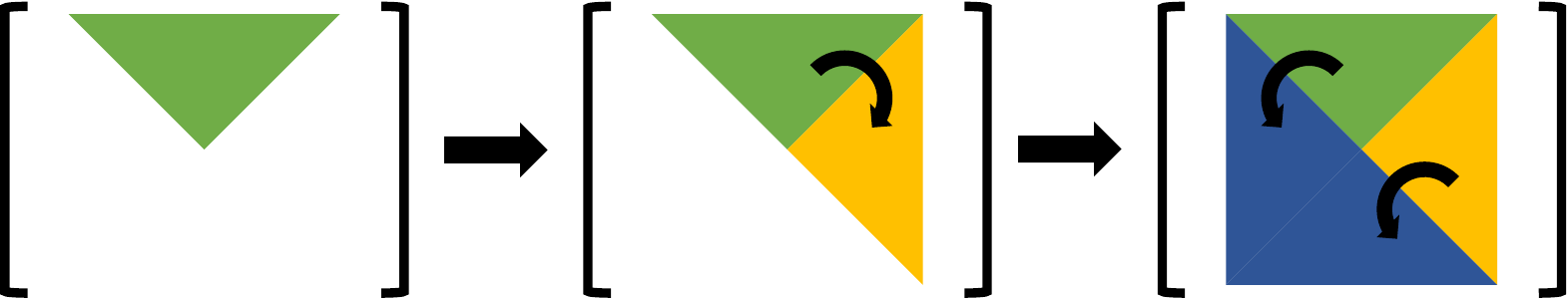}
    \caption{Illustration of the unfolding procedure for constructing a perfectly symmetric reconstructed vertex. We start by extracting the green rectangle from the reconstructed matrix and unfold it using the relations from Eq.~\eqref{eq:symm}}
    \label{fig:sym_schematic}
\end{figure}

\begin{figure}
    \centering
    \includegraphics[width=0.7\textwidth]{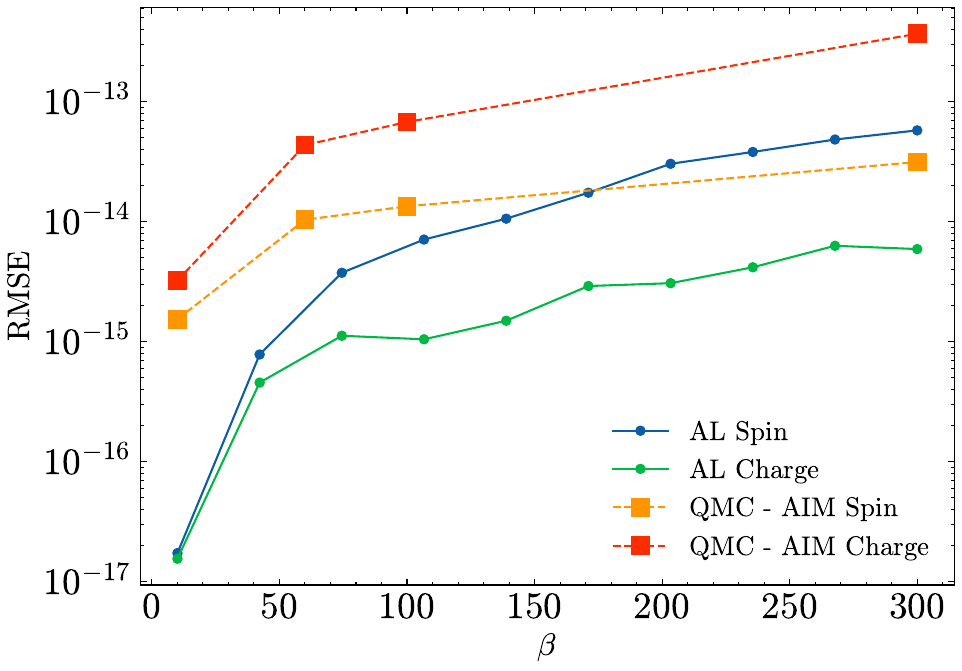}
    \caption{Comparison of the symmetry conservation in the wavelet decomposition with QMC. The RMSE between the reconstructed $\chi^{rec}$ and \emph{perfectly symmetric}, reconstructed $\chi^{rec}_{sym}$ is shown for the atomic limit at $U=1$ as a function of temperature at a decomposition level of $l=8$ and a threshold quantile of $q=0.99$. Here we chose a grid size of $512\times512$. The results are compared to the RMSE between the charge susceptibility of the AIM at $U=5.75$ and its symmetrized counterpart. The data for the AIM is obtained from Ref.~\cite{chalupa2021} Fig.1 therein.}
    \label{fig:qmc}
\end{figure}

\begin{figure}
    \centering
    \includegraphics[width=\textwidth]{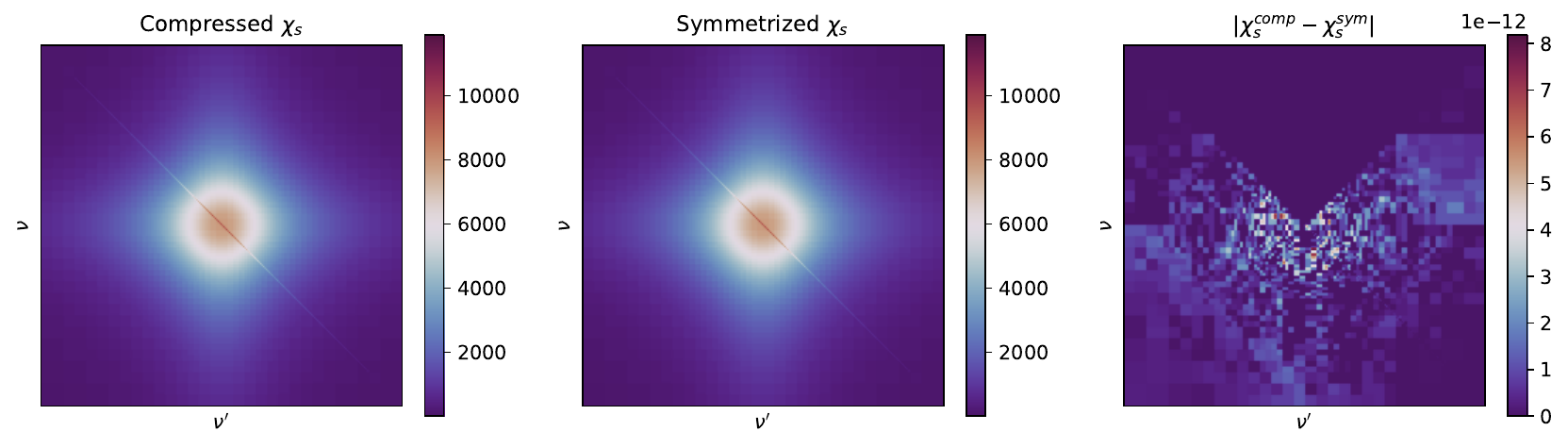}
    \caption{Example of the reconstructed $\chi^{rec}$ (left panel) and \emph{perfectly symmetric}, reconstructed $\chi^{rec}_{sym}$ (central panel) for the spin channel at $U=1$ and $\beta=1000$ alongside the absolute difference. For the wavelet decomposition a level of $l=8$ and a threshold of $q=0.99$ is chosen, whereas the frequency grid of the susceptibility has the size $512\times512$.}
    \label{fig:sym_compare}
\end{figure}

In addition to the overall compression performance of the DWT as discussed above, we would also like to address 
its ability to correctly capture the symmetries of the generalized susceptibilities. In the particular case of half filling considered here, the generalized susceptibilities can be shown to be real, symmetric and centrosymmetric matrices (for further details we refer to Refs.~\cite{rohringer2013phd, rohringer2012, rohringer2018, chalupa2018, springer2020}), which can be formulated via the following equation
\begin{equation}\label{eq:symm}
    \left(\chi^{\nu\nu'}\right)^* = \chi^{\nu'\nu} = \chi^{-\nu -\nu'}
\end{equation}
In order to analyze whether our wavelet-based compression method is capable of preserving these symmetries, we designed the following experiment. First the generalized susceptibility is compressed and reconstructed. Subsequently, we cut out a piece and use this to construct a perfectly symmetrized susceptibility consistent with Eq.~\eqref{eq:symm} as visualized in Fig.~\ref{fig:sym_schematic}. This newly obtained perfectly symmetric susceptibility $\chi^{rec}_{sym}$ is then compared to the original, reconstructed $\chi^{rec}$ to quantify the degree of symmetry violation without considering other reconstruction errors. In particular, we calculate the root mean squared error (RMSE) between these two quantities for the spin and charge susceptibilities with a frequency grid size corresponding to $512\times512$ at a decomposition level of $l=8$ and a threshold quantile of $q=0.99$, which for these parameters corresponds to the maximal possible compression. In addition, as a direct comparison we also perform a similar \emph{symmetry analysis} with two-particle quantities obtained via state-of-the-art quantum Monte Carlo (QMC) computations~\cite{w2dynamics}, where we compare the original Monte Carlo vertex with a symmetrized one. The motivation behind this comparison lies in the fact, that QMC provides a \emph{numerically exact} solution of the AIM. Hence, the degree of symmetry breaking can be controlled by choosing sufficiently high sampling statistics. Therefore, we can utilize high-quality QMC data as a benchmark to estimate to what degree symmetry violations occur in our wavelet-based compression scheme. To this end, we utilize data sets of the static, spin and charge susceptibilities of the Anderson impurity model (AIM) at different temperatures, namely $\beta=\{10, 60, 100, 300\}$, corresponding to the quantities from Fig. 1 in Ref.~\cite{chalupa2021} or equivalently from Ref.~\cite{chalupa2022phd}. The resulting RMSE values are presented in Fig.~\ref{fig:qmc}. We immediately observe a similar if not smaller RMSE for the wavelet-compressed susceptibilities compared to QMC, where interestingly for the latter the symmetries of the spin susceptibility seem to be less violated. In the case of the DWT on the other hand, consistent with our findings in Fig.~\ref{fig:cr_ssim}, the symmetry preservation of the charge as compared to the spin channel appears to be slightly better. In order to obtain a qualitative understanding of this symmetry analysis, Fig.~\ref{fig:sym_compare} provides an illustrative example of the proposed procedure. The left and central panels respectively show the reconstructed and perfectly symmetric susceptibilities for $\beta=1000$ and $U=1$ at the same compression parameters as used for the results in Fig.~\ref{fig:qmc}, whereas in the right panel the absolute difference between these two is plotted. One can indeed observe that apart from the triangular piece that was cut out to symmetrize the susceptibility the error is merely numerical noise. Lastly, it is noteworthy that other bases than the \emph{Haar} wavelet prove to be significantly less suitable for correctly capturing the symmetries of the half-filled atomic limit susceptibilities (see Appendix \ref{sec:app_symm}).

\subsection{Reconstruction of physical quantities}
\subsubsection{Compression of the physical susceptibilities}\label{sec:cs_sep}
\begin{figure}[h!]
    \centering
    \includegraphics[width=0.8\textwidth]{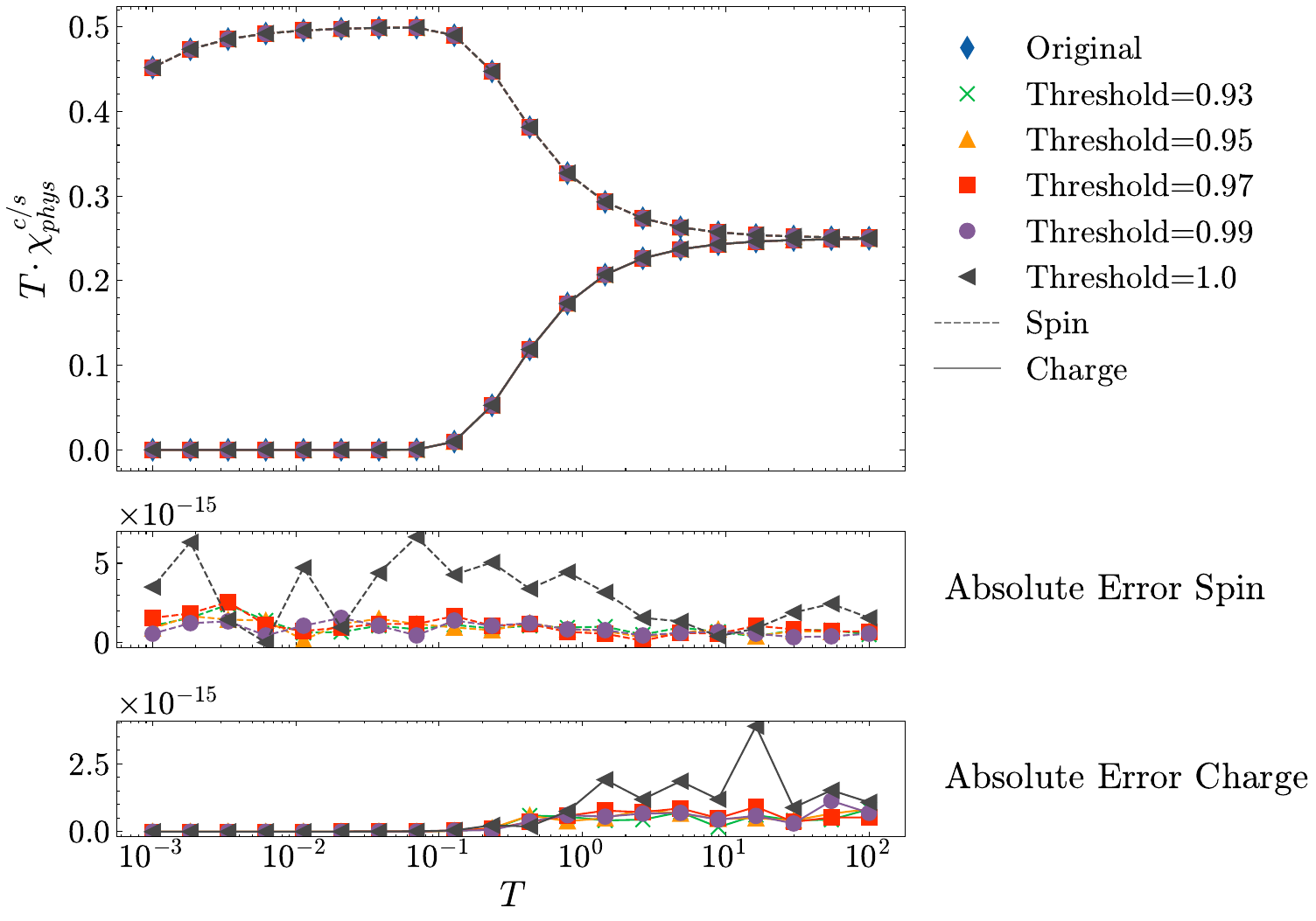}
    \caption{Physical spin and charge susceptibilities $\chi^{c/s}(T)$ as computed from the respective generalized quantities for various threshold quantiles. Here, a decomposition level of $l=10$ and a frequency grid with a size of $2048\times2048$ is used. In the lower panels, the absolute errors with respect to the original uncompressed physical susceptibilities are displayed.}
    \label{fig:cs_sep}
\end{figure}

In the current section, we focus on characteristic physical properties or observables associated to the atomic limit susceptibilities and how these are affected by our proposed wavelet-based compression algorithm. As a first step, we calculate the physical response functions $\chi_{s/c}$ from the generalized spin and charge susceptibilities. These are then compared to the physical susceptibilities obtained from the respective compressed quantities. We are particularly interested in the temperature behaviour of $T \cdot \chi_{s/c}$. Therefore, we perform the wavelet compression for various threshold quantiles to the generalized spin and charge susceptibilities at different temperature values. In this case we chose the maximal possible decomposition level of $l=10$ for a frequency grid size of $2048\times2048$. The corresponding results are shown in Fig.~\ref{fig:cs_sep}. The upper panel shows the comparison between the true, uncompressed and the compressed quantities, whereas in the lower two panels the absolute errors for the spin and charge channel are displayed. We immediately observe a perfect reconstruction of the physical response functions even for the maximal threshold quantile of $q=1$, where essentially only the final, single approximation coefficient is kept. The fact, that even after such a drastic truncation a perfect reconstruction can be achieved, can be explained in the following way. The scaling function of the \emph{Haar} wavelet, responsible for capturing the low frequency dynamics of the signal, computes the average between two neighbouring pixels or data points. Recursively applying this scaling function to the signal at every level and subsequently reconstructing it after truncating all detail coefficients will hence result in a new signal where each point is given by the mean value of the original signal. Thus, performing the Matsubara summation to obtain the physical susceptibilities will then, even for a maximal threshold of $q=1$, lead to the same result if one would perform the summation over the original uncompressed data. We also want to stress, on an unrelated note, that the failure of the spin susceptibility to converge to a value of $0.5$ in the zero temperature limit is due to the fact, that we perform the Matsubara summation over a finite frequency box and do not consider the appropriate asymptotics of the generalized susceptibilities.

\subsubsection{Eigenvalues and eigenvectors of generalized susceptibilities}\label{sec:evs}

As mentioned in the Introduction, in spite of its simple nature, the atomic limit of the Hubbard model is capable of capturing non trivial features of strongly correlated electron systems in the strong coupling limit. An arguably important aspect is the breakdown of the self-consistent perturbation theory, as well as its formal \cite{schafer2013divergent, kozik2015nonexistence} and physical \cite{reitner2020attractive,chalupa2021fingerprints} manifestations. We recall that such a breakdown is signalled by the divergences of the two-particle irreducible vertex functions, which are related to the generalized susceptibilities via the Bethe-Salpeter equation (BSE) \cite{rohringer2012, rohringer2013phd, rohringer2018, thunstrom2018analytical}. Specifically, for certain parameter sets, some eigenvalues of the generalized susceptibility in the charge channel vanish making its matrix representation singular and the corresponding BSE non invertible. The first vanishing eigenvalue, associated to the first divergence of the irreducible vertex, marks thus the onset of the non-perturbative regime. 
Evidently, the eigenvalues as well as the corresponding eigenvectors' structure encode \cite{gunnarsson2017breakdown,springer2020interplay,reitner2020attractive,vanloon2020bethesalpeter, reitner2023protection, kowalski2023thermodynamic} significant information for the physics of  strongly correlated many-electron systems.

For this reason, in the following numerical experiments, we %would like to 
examine the effect of the DWT on the eigenvalue and eigenvector structure of the generalized susceptibilities. To begin with, we will analyze the capabilities of the proposed wavelet-based compression to locate the exact point in the Hubbard atom's $U-T$ phase-diagram where the irreducible vertex function in the charge channel diverges or, equivalently, where the eigenvalues of the generalized charge susceptibility become zero \cite{schafer2016nonperturbative,thunstrom2018analytical}. 
%in correspondence of the zeros of the eigenvalues of the local charge susceptibilities

\begin{figure}
    \centering
    \includegraphics[width=0.6\textwidth]{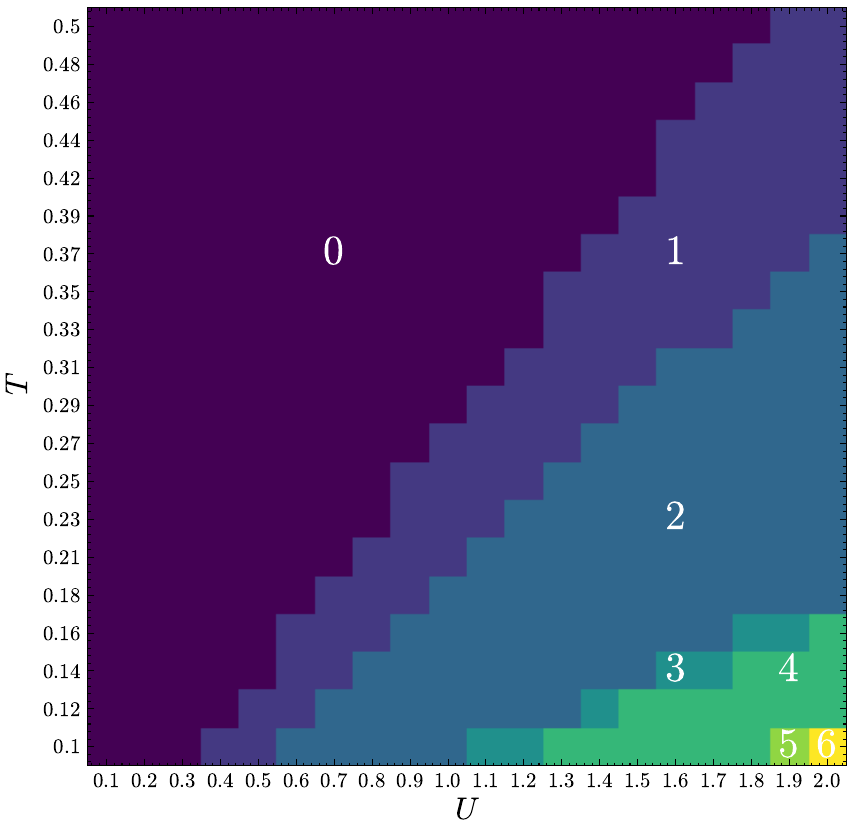}
    \caption{$U-T-$phase-diagram for the Hubbard atom at half-filling showing the number of negative eigenvalues of the generalized charge susceptibility. The resulting lines hence denote the divergence lines of the irreducible charge vertex of the Hubbard atomic limit \cite{schafer2016nonperturbative}.}
    \label{fig:div_lines}
\end{figure}

Fig.~\ref{fig:div_lines} plots the number of negative eigenvalues of the charge susceptibility as a function of $U$ and $T$, where for the sake of simplicity and illustrative purposes we chose a very coarse grid. The emerging transitions, hence mark the divergence lines where the eigenvalues of the charge susceptibility vanish. In order to investigate whether the DWT is able to determine the location of the vertex divergence, we fix the temperature to $\beta=1000$ and find via binary search the interaction value where the lowest eigenvalue becomes zero. The eigenvalues are computed for the true, uncompressed as well as the compressed charge susceptibility at various threshold quantiles and a decomposition level of $l=6$. In Tab.~\ref{tab:div1} we report the resulting values of $U$ for which the first eigenvalue vanishes, where the error indicates the precision we set for the binary search algorithm. Additionally, for the sake of completeness, we also provide the relative compression as well as the SSIM score. Unsurprisingly, in this case, where we consider the charge susceptibility which showed very high and robust compression metrics in Sec.~\ref{sec:compr}, we observe no deviation from the true critical $U$-value up to our set precision. 

Since, the compression in the charge channel works almost perfectly, it is perhaps more instructive to consider the spin susceptibility as well which for sufficiently low temperatures leads to a significant deterioration of the SSIM scores. Therefore, in a similar fashion as before, we will now investigate the eigenvalues of the generalized spin susceptibility at $\beta=1000$ and $U=1$. Specifically, we will monitor how the maximal positive eigenvalue, which is, to a first approximation, responsible for the Curie behaviour of the static spin response of the system \cite{springer2020interplay}, behaves upon compression with varying threshold quantiles. Here again, we have chosen a decomposition level of $l=6$ and a frequency grid with the size $512\times512$. The results are presented in Tab.~\ref{tab:lamb_max}, where in addition to the maximal eigenvalue $\lambda_{max}$, we also report the eigenvalue scaled by its corresponding spectral weight $w_{max}$, which is defined as $w_{max} = \sum_{\nu} V^{-1}_{max}(\nu) \times \sum_{\nu'} V_{max}(\nu')$. As before, we also show the relative compression ratio as well as the SSIM score. In the case of the spin channel, we now observe a slight deviation of the quantities under consideration from the true, uncompressed values. Nonetheless, up to a relative error in the order of magnitude of $~\mathcal{O}(10^{-4})$, $\lambda_{max}$ and $w_{max}$ are hardly affected by the compression, despite the relatively poor SSIM score. 

\begin{figure}
    \centering
    \includegraphics[width=0.9\textwidth]{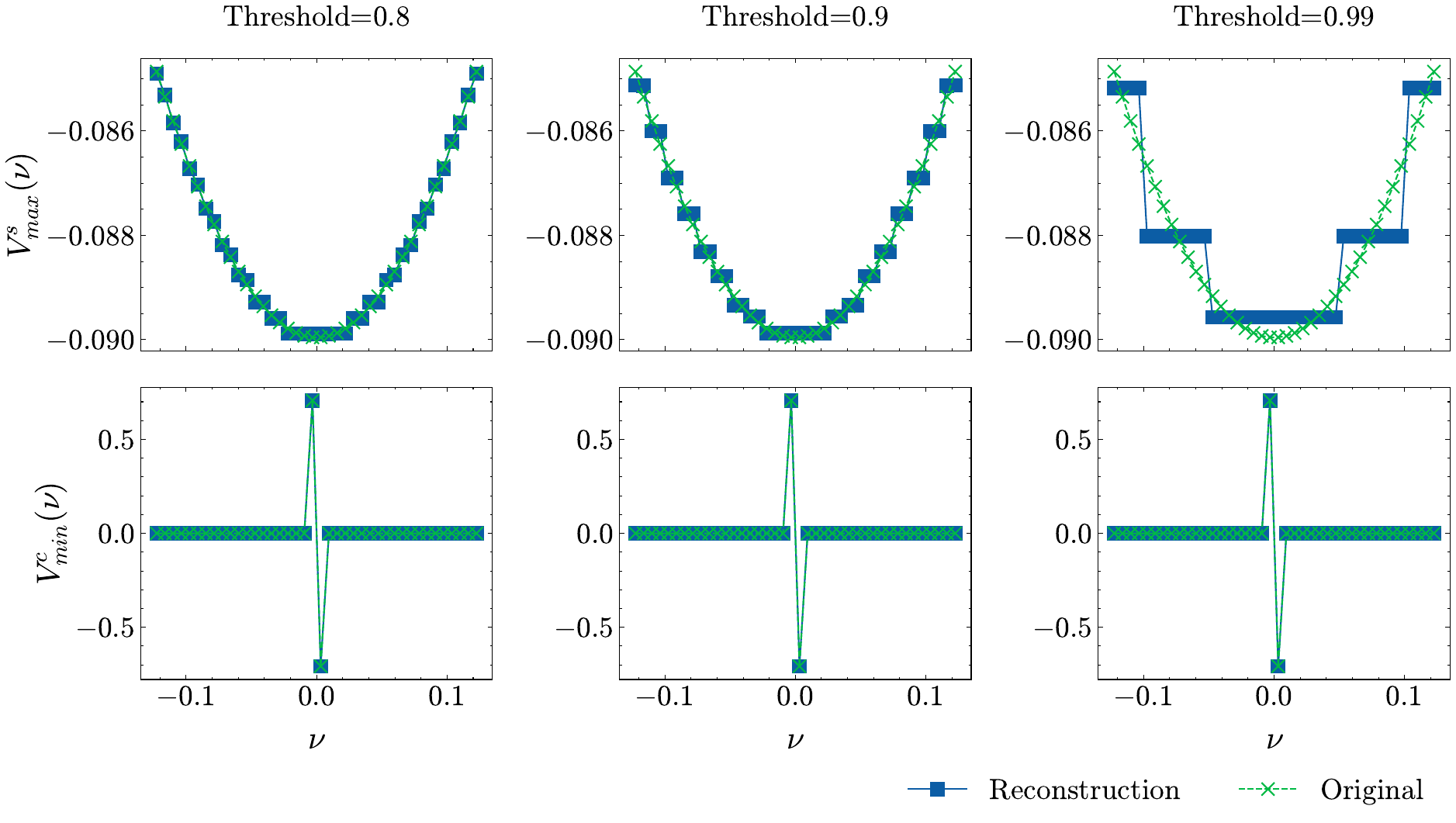}
    \caption{Upper panels: Eigenvector corresponding to the maximal eigenvalue of the reconstructed generalized spin susceptibility for $U=1$ and $\beta=1000$ at different threshold quantiles compared to the eigenvector of the original, uncompressed susceptibility. Lower panels: Eigenvector corresponding to the lowest eigenvalue for the reconstructed charge susceptibility at various threshold quantiles compared to the eigenvector calculated from the uncompressed quantity. For the decomposition level $l=6$ and a frequency grid size of $512\times512$ were chosen. Here, the eigenvectors were computed at $U=0.00362759873$ and $\beta=1000$, i.e. where the lowest eigenvalue vanishes.}
    \label{fig:v1}
\end{figure}

\begin{figure}
    \centering
    \includegraphics[width=0.9\textwidth]{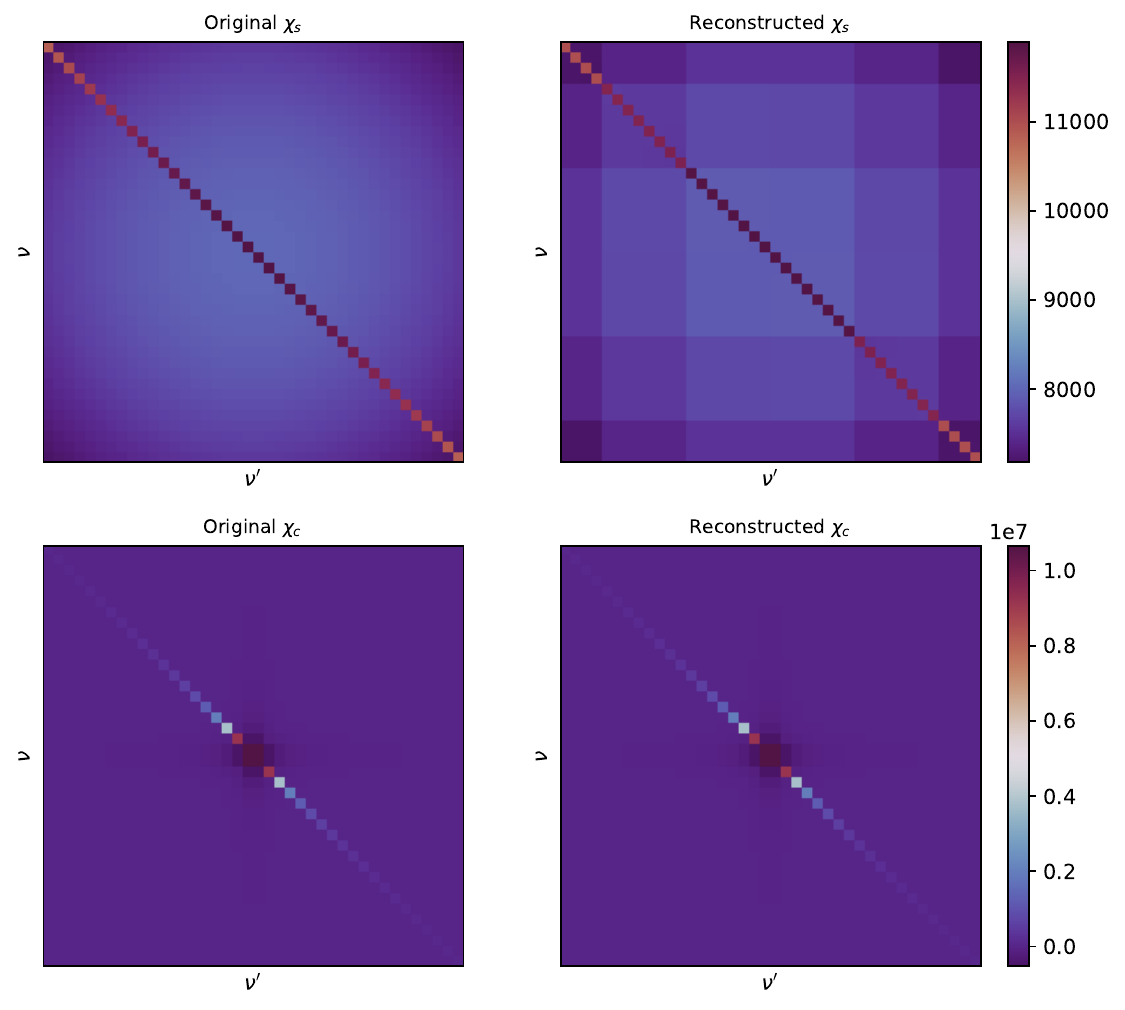}
    \caption{Zoom on the first 20 Matsubara frequencies of the charge and spin susceptibilities alongside their respective reconstructions from the parameter sets of Fig.~\ref{fig:v1}, where the threshold quantile is set to $q=0.99$.}
    \label{fig:compr_zoom}
\end{figure}

Furthermore, to also obtain an insight into the \emph{structural} behaviour of the generalized susceptibilities with varying compression strengths, we have also investigated in detail how the associated eigenvectors for the parameter sets in the two cases discussed above are affected by the DWT. In particular, we looked  (i) at the eigenvector corresponding to the lowest eigenvalue of the charge susceptibility at $U=0.003628$, where said eigenvalue becomes zero, as well as (ii) the eigenvector of the largest eigenvalue of the spin susceptibility at $U=1$, because the first is associated to the breakdown of the perturbation expansion and the second controls the Curie magnetic response of the system. 
In both cases we fixed the temperature to $\beta=1000$. We compare the eigenvectors which we obtained from the compressed susceptibilities to the original ones for various threshold quantiles and a decomposition level of $l=6$, whereas the frequency grid size was again set to $512\times512$. The results are shown in Fig.~\ref{fig:v1}, where the original and compressed eigenvectors of the spin and charge susceptibilities are displayed in the plots of the upper and lower row respectively. In contrast to the charge channel, whose lowest eigenvector, which is fully localized in frequency, is always perfectly reconstructed, we can observe an increasingly coarser reconstruction of the broad structure of the largest eigenvector of the spin susceptibility. This different behavior can be qualitatively explained upon inspecting the generalized susceptibilities in Matsubara frequency space more closely. In Fig.~\ref{fig:compr_zoom} we show, for the same parameter sets as before, the original spin and charge susceptibilities and their compressed counterparts, where the threshold quantile is now set to $q=0.99$. One can easily note, then, that the generalized spin susceptibility is much broader distributed in frequency space than its charge counterpart, which exhibits a frequency behaviour almost reminiscent of a delta-function. These characteristics are also very distinctively manifested in the frequency structure of the respective eigenvectors we have considered. As a consequence, the wavelet transform is not able to capture all the information contained in the spin susceptibility with fewer wavelet coefficients as opposed to the charge channel, which due to its sharper features requires far less coefficients. This leads to the coarse-graining effect observed in the reconstructed eigenvectors (upper row in Fig.~\ref{fig:v1}) as well as the compressed spin susceptibility (upper right panel in Fig.~\ref{fig:compr_zoom}) when too many detail coefficients are truncated. We also want to highlight, that this aspect gives a reasonable explanation as to why in Fig.~\ref{fig:cr_ssim} a significantly smaller SSIM score was observed for the spin susceptibility at higher $\beta$ values.

\begin{table}[htbp]
\centering
\begin{tabular}{cccc}
\hline
Threshold Quantile  & Transition in $U$       & Memory Compr. & SSIM \\ \hline
0                  & $0.003628 \pm 10^{-7}$ & -                & -    \\ 
0.91               & $0.003628 \pm 10^{-7}$ & 0.82             & 1.0  \\
0.93               & $0.003628 \pm 10^{-7}$ & 0.86             & 1.0  \\
0.95               & $0.003628 \pm 10^{-7}$ & 0.90             & 1.0  \\
0.97               & $0.003628 \pm 10^{-7}$ & 0.94             & 0.999  \\
0.99               & $0.003628 \pm 10^{-7}$ & 0.98             & 0.901  \\ \hline
\end{tabular}
\caption{Value of the Hubbard $U$ where the lowest eigenvalue of the reconstructed generalized charge susceptibility $\chi^c$ at $\beta=1000$ vanishes, for a decomposition level of $l=6$ and a frequency size of $512\times512$ for various threshold quantiles. Additionally, the compression metric for the memory and the SSIM are reported. The divergence in the original, uncompressed vertex is also found to be at $U = 0.003628 \pm 10^{-7}$.}
\label{tab:div1}
\end{table}

\begin{table}[htbp]
\centering
\begin{tabular}{ccccc}
\hline
Threshold Quantile  & Value of $\lambda_{max}$ & $\lambda_{max} \times w_{max}$ & Memory Compr. & SSIM \\ \hline
0                  &  987605.434  & 326.211 & - & - \\
0.91               & $987577.921$ & 326.203 & 0.82             & 0.857  \\
0.93               & $987569.835$ & 326.202 & 0.86             & 0.835  \\
0.95               & $987553.718$ & 326.198 & 0.90             & 0.790  \\
0.97               & $987503.102$ & 326.184 & 0.94             & 0.699  \\
0.99               & $987246.247$ & 326.117 & 0.98             & 0.525  \\ \hline
\end{tabular}
\caption{Value of the maximal eigenvalue of the generalized charge susceptibility at $U=1$ and $\beta=1000$ vanishes, for a decomposition level of $l=6$ and a frequency size of $512\times512$ for various threshold quantiles. Additionally, the compression metric for the memory and the SSIM are reported. The magnitude of the maximal eigenvalue of the original, uncompressed susceptibility is $\lambda_{max} = 987605.434$ and factoring the spectral weight gives $\lambda_{max} \, w_{max} = 326.211$.}
\label{tab:lamb_max}
\end{table}

\subsubsection{Schwinger-Dyson-Equation with the compressed $G^{(2)}$}

\begin{figure}
    \centering
    \includegraphics[width=0.95\textwidth]{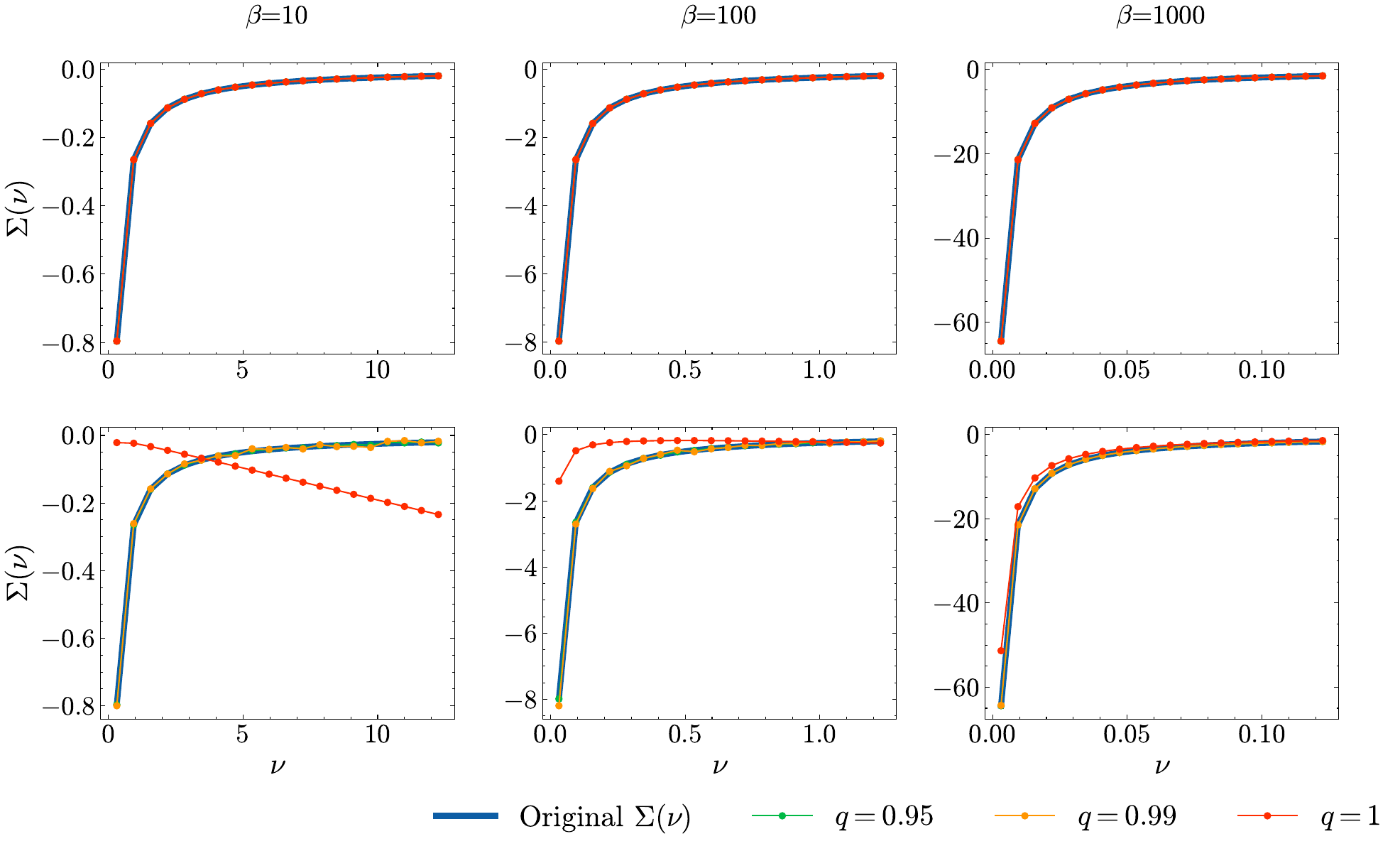}
    \caption{The self-energy computed from the SDE  with the original, uncompressed two-particle Green's function $G^{(2)}(\nu, \nu', \omega)$ compared to the one obtained from a compressed two-particle Green's function at various threshold quantiles, for different temperatures. Upper panel: The compression is performed on $(\nu', \omega)$-slices along the $\nu$-frequency range, where no subsequent Matsubara summation is performed in the SDE. Lower panel: Here the compression is performed on the entire two-particle Green's function as a function of three frequencies.}
    \label{fig:se_haar}
\end{figure}

The final set of numerical experiments will be devoted to the Schwinger-Dyson equation (SDE), also referred to as the equation of motion for the one-particle Green's function. Our goal is to analyze the effect of the wavelet-based compression of the two-particle Green's function, the central component of the SDE, on the resulting self-energy obtained from the equation of motion. We note, that in this case the Green's function depends on three frequencies, which will require a three-dimensional DWT. The SDE for the local self-energy is given by
\begin{equation}
    \Sigma(\nu) = \frac{Un}{2} - \frac{U}{\beta^2}\sum_{\nu', \omega} F^{\nu \nu' \omega}_{\uparrow \downarrow} \, G(\nu') \, G(\nu'+\omega) \, G(\nu+\omega),
\end{equation}
where $F_{\uparrow\downarrow}$ denotes the corresponding spin components of the full vertex function \cite{rohringer2013phd}. The above equation can be equivalently formulated in terms of the two-particle Green's function \cite{rohringer2013phd}
\begin{equation}
    \Sigma(\nu) = \frac{Un}{2} - \frac{U}{\beta^2}\sum_{\nu', \omega} G^{-1}(\nu) \, G^{(2), \, \nu\nu'\omega}_{\uparrow \downarrow}.
\end{equation}
In order to assess the effect of the DWT on the self-energy, we will extract it from the SDE using the analytic two-particle Green's function of the Hubbard atom and compare the results with the self-energy obtained from a wavelet-compressed $G^{(2)}$. The compression will be done at various threshold quantiles and a maximal decomposition level for a frequency grid of $64\times64\times64$. The original and reconstructed self-energies for various temperature values are reported in the lower panel of Fig.~\ref{fig:se_haar}. In this case, we employed a three-dimensional DWT to compress the entire two-particle Green's function before plugging it back into the SDE. We generally observe larger reconstruction errors for smaller values of $\beta$, this is especially noticeable for larger threshold quantiles such as $q=1$. Upon lowering the temperature, these errors are also further reduced. This, at first glance, anomalous behaviour can be explained by the analytic properties of $G^{(2), \, \nu\nu'\omega}_{\uparrow \downarrow}$. The two-particle Green's function under consideration follows a Curie-Weiss law and shows a $1/T$ divergence along the main diagonal $\nu=\nu'$ for $\beta \rightarrow \infty$ (for the analytic expression of $F^{\nu\nu'\omega}_{\uparrow \downarrow}$ we refer to Eq. (2.223) in Ref.~\cite{rohringer2013phd}). This leads to very sharp features along the main diagonal compared to the background $\nu \neq \nu'$, which, as we discussed before in Sec.~\ref{sec:evs} when we considered the eigenvectors of the generalized susceptibilities, are significantly easier for the DWT to capture accurately with fewer coefficients. 

Evidently, exploiting the knowledge we gained from compressing the physical susceptibilities in Sec.~\ref{sec:cs_sep}, it is also possible to perform the compression for two-dimensional slices along the $\nu$-frequency axis. This way, as subsequently in the SDE a Matsubara summation in the $\nu'$ and $\omega$ frequency direction will be performed, we can rest 
assured that the wavelet-based compression will yield \emph{exact} results. In the upper panels of Fig.~\ref{fig:se_haar} the reconstructed self-energies are compressed in this manner. And indeed, we observe a perfect reconstruction even for a threshold quantile of $q=1$, consistent with our findings from Fig.~\ref{fig:cs_sep}. Needless to say, this way of performing the wavelet compression, might pose higher memory requirements than simply performing a three-dimensional DWT for the entire two-particle Green's function.

\section{Conclusions and outlook}

This study has shown that wavelet compression methods can efficiently represent generalized susceptibilities and self-energies within the Hubbard model at half-filling in its atomic limit. We have shown that the wavelet method can efficiently address computational challenges associated with storing and manipulating the non-zero temperature correlation functions in systems characterized by strong interactions, and have discussed how to characterize the efficiency of the compression. The results demonstrate that the wavelet method is versatile and resource-efficient. As a concluding remark, we can give a crude estimation on the storage capacities of the wavelet transform when it comes to two-particle quantities often encountered in many-body theory. In the general case, a two-particle correlation function will depend on three frequency and three momentum indices in addition to possibly multiple orbital degrees of freedom. Depending on the dimensionality of the lattice considered for the Hamiltonian the overall memory scaling will be proportional to $\mathcal{O}(N_{orb}^4N_{\nu}^3N_{k}^{3d})$ with $N_{orb}$ orbitals, $N_{\nu}$ number of Matsubara frequencies for each of the three frequency indices and $N_{k}$ datapoints for the momentum grids in each lattice dimension. One can immediately see that this scaling will quickly become immensely challenging to handle. For instance, neglecting, for the sake of simplicity, the orbital and momentum dependence and considering only $N_{\nu}=1000$ for all three Matsubara frequencies, the memory requirements to store $N_{\nu}^3$ datapoints as double precision floating points will already be of the order of 8Gbytes. In Sec.~\ref{sec:results} we observed that the compression ratios where the reconstruction quality was still considerably good were between 90 \% and 98 \%, which would compress the considered two-particle object down to 160 - 800Mbytes yielding a significant reduction of memory requirements.

The successful representation of the atomic limit opens the door to the broader usage of signal processing and data compression principles. Future directions include the adaptation of wavelet-based compression techniques to explore non-zero doping conditions, moving away from the symmetry constraints of half filling. Additionally, future studies will aim to extend its applicability to more realistic scenarios, such as the investigation of two-particle Green's functions within the Anderson impurity model or the Hubbard model at intermediate or weak coupling. A distinctive feature of the wavelet-based approach, distinguishing it from the previously mentioned IR and DLR compression methods, is its capacity for a data-agnostic application. This characteristic allows for its use in both momentum-dependent and real-frequencies-dependent Green's functions. In this respect, particularly intriguing will be its potential application in addressing interacting systems under nonequilibrium conditions, thereby broadening the scope of research possibilities in the field. 

We observe that a seamless integration of wavelet methods into the existing workflow of numerical methods for the many-body problem requires the application of fundamental mathematical operations inherent to many-body physics — specifically, convolutions and inversions — within the wavelet basis~\cite{perezrendon2004, drori2003, harten1994}. Achieving this understanding represents a critical step to establish the wavelet-based framework as an integral component of numerical methodologies for tackling complex many-body systems.

From a pure data science perspective, research directions may extend toward broadening the applicability of our proposed compression scheme to higher-dimensional wavelet functions, ones that go beyond mere products of one-dimensional functions and instead possess intrinsic higher-dimensional characteristics. Pursuing this research trajectory holds the potential to provide even more effective compression techniques, enhancing the overall efficiency. An alternative approach involves leveraging machine learning techniques to obtain optimal performance in the compression process. This can be achieved by incorporating learnable thresholding functions and employing dictionary learning to determine the set of basis functions that most effectively represent the underlying data~\cite{zhuang1994, recoskie2018, michau2022}.

In conclusion, this work serves as a pioneering proof-of-principle application of wavelet compression in the field of many-body physics, opening new possibilities for future research efforts and providing a valuable contribution to the ongoing evolution of quantum many-body physics methodologies.

\section*{Acknowledgments}

The authors thank Markus Wallerberger for valuable discussions. %and ... for carefully reading the manuscript. 
We would also like to thank Patrick Chalupa-Gantner for providing the QMC data of the AIM as well as Matthias Reitner for sharing his code to calculate the atomic limit quantities. We acknowledge financial support from the Deutsche Forschungsgemeinschaft (DFG) within the research unit FOR5413 (Grant No. 465199066) (S.A.) and from the Austrian Science Fund (FWF) (grant-DOI 10.55776/I 5487 (A.T.) as well as grant-DOI 10.55776/I 5868 (Project P01, part of the FOR 5249 [QUAST] of the German Science Foundation, DFG) (E.M.)). The research leading to these results has received funding from the European Union’s Horizon 2020 research and innovation programme under the Marie Sk{\l}odowska-Curie Grant Agreement No. 897276 (D.D.S.). J.Z. and A.J.M acknowledge support from the NSF MRSEC program through the Center for Precision-Assembled Quantum Materials (PAQM) - DMR-2011738. The Flatiron Institute is a division of the Simons Foundation.

\section*{Data Availability Statement}
The manuscript has associated data in a data repository. [Author's comment: A data set containing all numerical data and plot scripts used to gnerate the figures of this publication is publicly available on the TU Wien Research Data repository \cite{data_repo}.]

\begin{appendices}    

\section{Effect of different frequency grid sizes and wavelet bases}\label{sec:app_gridsize}

\begin{figure}
    \centering
    \includegraphics[width=0.95\textwidth]{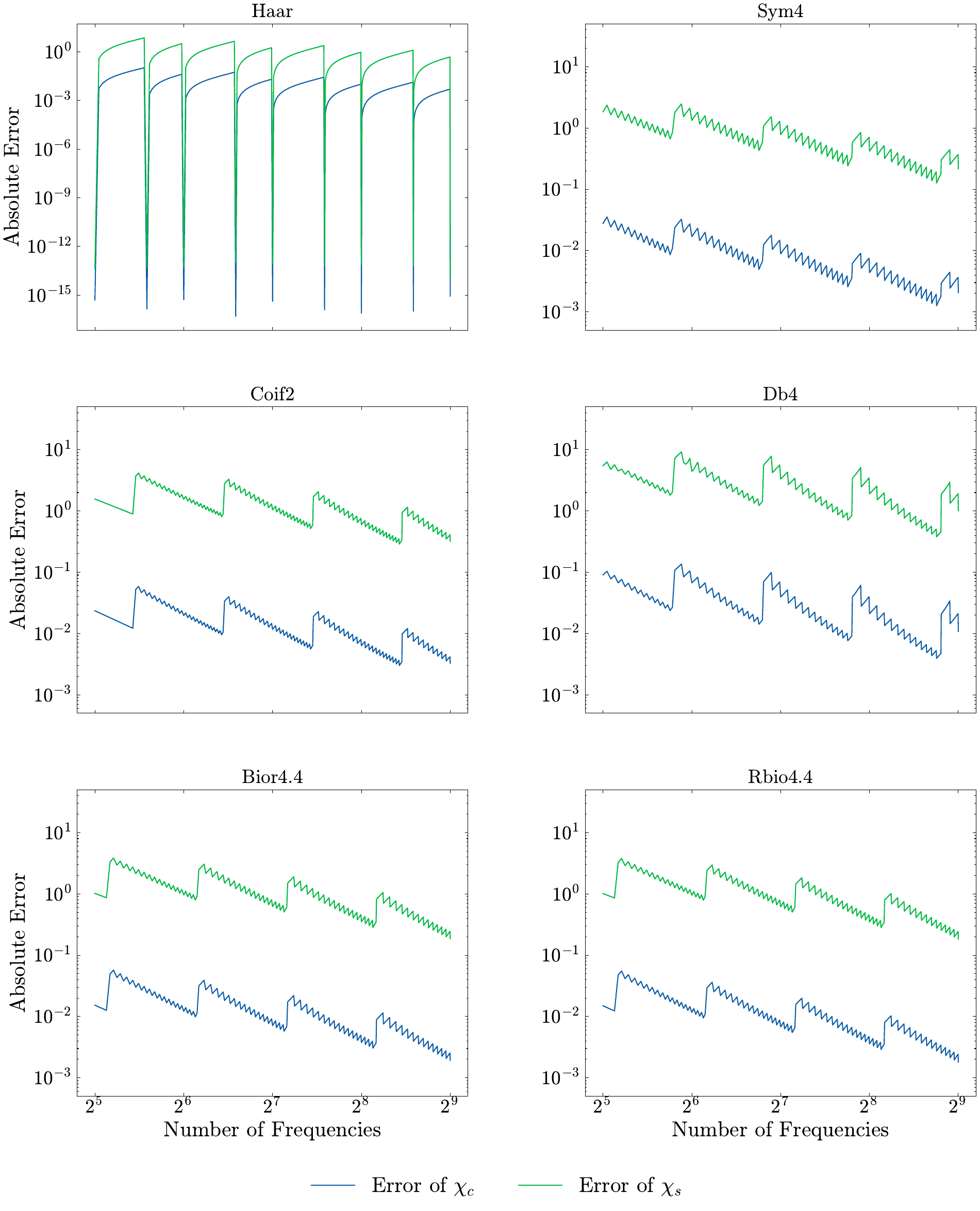}
    \caption{Absolute errors between the original and the reconstructed data as a function of the grid size, for different wavelet bases. The calculations are performed at $U = 1$ and $\beta = 100$, with a threshold quantile of $q=1$. The decomposition level is chosen to be the maximum allowed level for a given vertex size.}
    \label{fig:wl_basis}
\end{figure}

We discuss here the dependence on the frequency grid as well as the effects induced by the wavelet basis choice, for the data shown in Fig. \ref{fig:cs_sep}. In addition, we provide results for different families of wavelets and how their use influences the quality of the compression and reconstruction.
For this, we ran the compression algorithm with a threshold quantile of $q=1$, meaning that only the low frequency approximation will be retained. The number of frequencies per dimension of a vertex and thus the size of it will be varied from $2^5$ up to $2^{9}$. For each vertex size, we set the decomposition level to its maximum allowed value, i.e. to the deepest possible level, where the length of the signal to be decomposed is equal or larger than the filter length of a given wavelet \cite{Lee2019}.
%\EM{i.e. to the deepest level of decomposition a specific vertex size can support without being compromised by edge effects. These are for instance caused by the fact that, for the maximum level decomposition, the signal length must be larger than the filter length of the given wavelet \cite{Lee2019}.} 
We then compute the absolute error between the reconstructed and the original physical susceptibility for both the spin and the charge channel and plot it over the frequency size in Fig. \ref{fig:wl_basis}. This plot shows the results for different wavelet families. We observe a very interesting pattern for the Haar wavelet where the relative error drops down to approximately $~10^{-14}$ representing a perfect reconstruction at vertex sizes that are powers of two as well as consecutive sums of powers of two. Since the Haar wavelet's filter length of $2$ allows for a decomposition exactly halving the approximation in each step, the powers of two can be decomposed without any significant loss of information. A similar property arises from a sum of two consecutive powers of two represented by:
\begin{equation}
2^n+2^{n+1}=2^n+2 \cdot 2^n=2^n(1+2)=2^n \cdot 3
\end{equation}
where up to a %remaining 
factor of $3$, the halving of the grid size can be applied repeatedly as well.

All other wavelet families exhibit similar qualitative results. We observe jumps in the relative error at those points where the decomposition level can be increased by one resulting in a pruning of more coefficients and thus a coarser approximation after the reconstruction. Between those jumps, we observe a downwards trend in the absolute error which can be traced back to the larger vertices that retain more information in the approximation coefficients leading to a better reconstruction.

\section{Symmetry preservation in different wavelet bases}\label{sec:app_symm}

\begin{figure}
    \centering
    \includegraphics[width=0.9\textwidth]{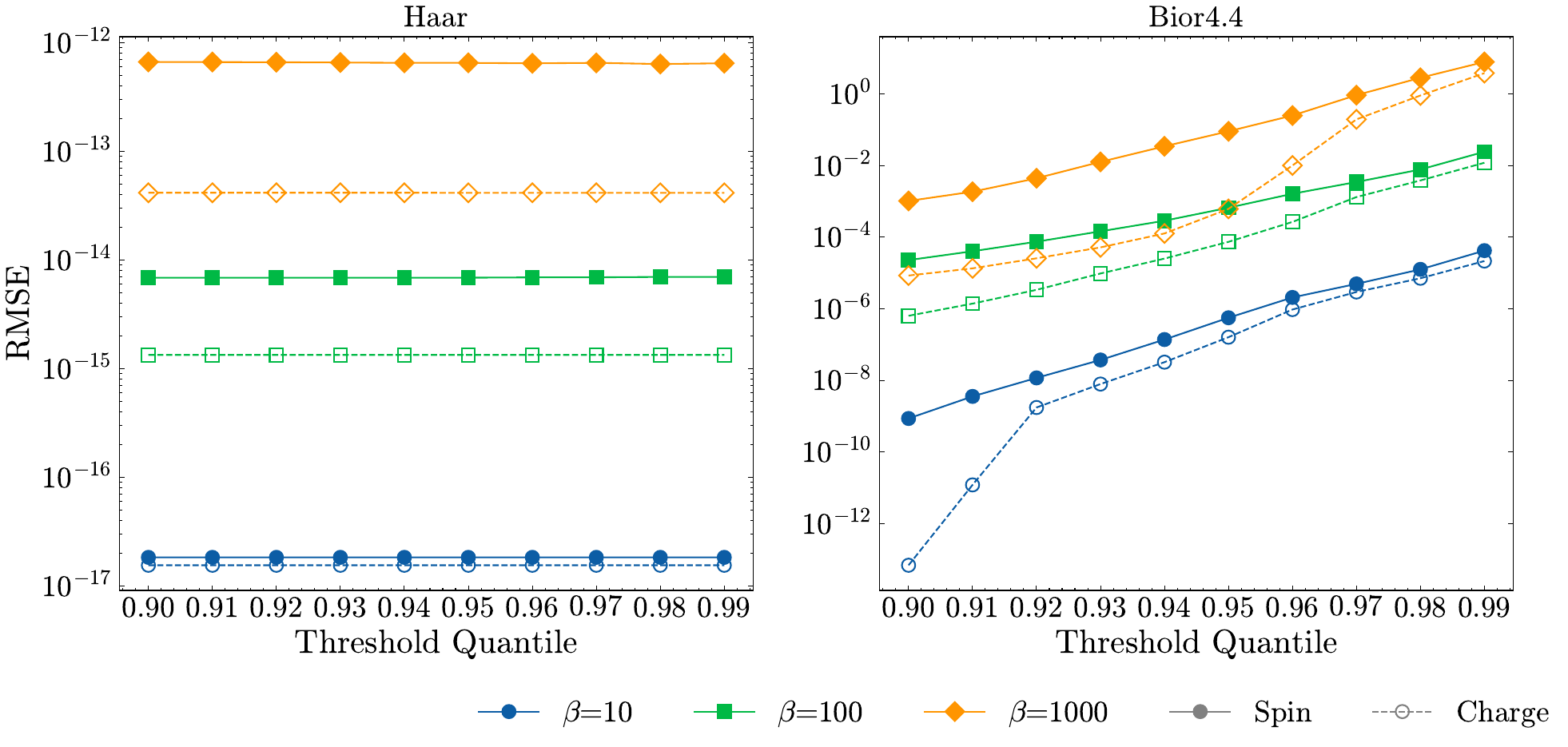}
    \caption{RMSE for the generalized spin (bold lines) and charge (dashed lines) susceptibilities at $U=1$ and $\beta=\{10,100,1000\}$ computed for the reconstructed $\chi^{rec}$ and \emph{perfectly symmetric}, reconstructed $\chi^{rec}_{sym}$. The decomposition level is $l=6$ and the size of the frequency grid $512\times512$. The left panel shows the reconstruction error for \emph{Haar} wavelets whereas in the right panel the biorthogonal wavelet basis \emph{bior4.4} was utilized.}
    \label{fig:sym_sus}
\end{figure}

\begin{figure}
    \centering
    \includegraphics[width=\textwidth]{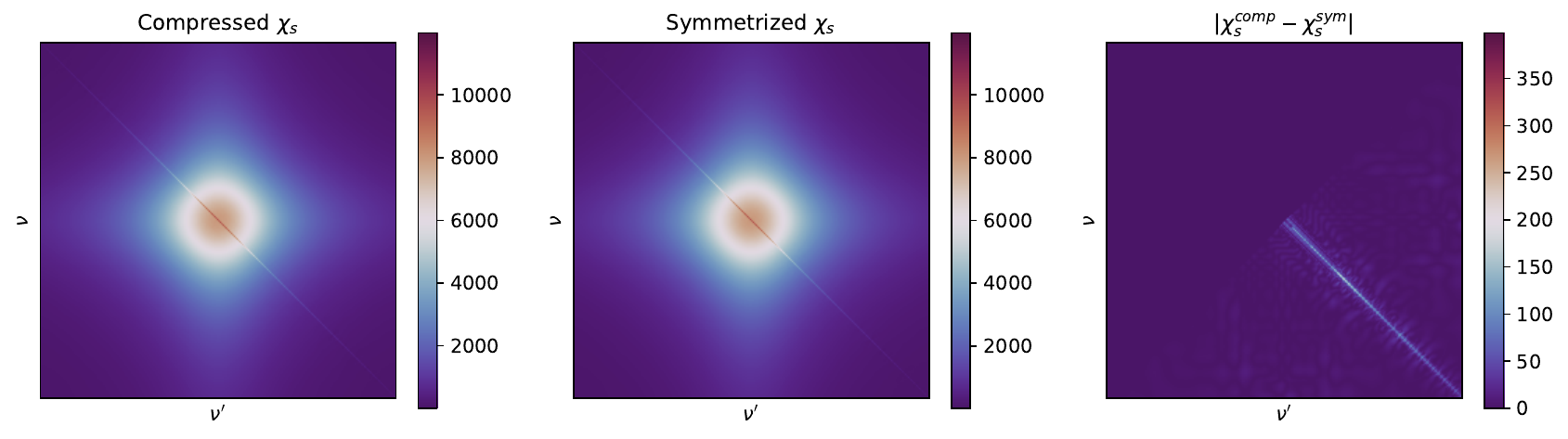}
    \caption{Example of the reconstructed $\chi^{rec}$ (left panel) and \emph{perfectly symmetric}, reconstructed $\chi^{rec}_{sym}$ (central panel) for the spin channel at $U=1$, alongside the absolute difference. For the wavelet decomposition a level of $l=6$ and a threshold of $q=0.99$ is chosen, whereas the frequency grid of the susceptibility has the size $512\times512$. In this particular case, in contrast to Fig.~\ref{fig:sym_compare} the biorthogonal wavelet basis \emph{bior4.4} was employed for the compression.}
    \label{fig:sym_compare_bior}
\end{figure}

In this section we will briefly discuss how other wavelet bases compare to \emph{Haar} wavelets in terms of the symmetry preservation discussed in Sec.~\ref{sec:sym}. Although we do the comparison here only for biorthogonal wavelets, other bases yield qualitatively similar results.

In close analogy to the analysis in Sec.~\ref{sec:sym}, we compute the RMSE between a compressed and reconstructed susceptibility $\chi^{rec}$ and its perfectly symmetrized counterpart $\chi^{rec}_{sym}$ once using \emph{Haar} wavelets and once the biorthogonal wavelet basis \emph{bior4.4}. The compression at a maximal possible decomposition level of $l=8$ and $l=4$ for  \emph{Haar} and biorthogonal wavelets respectively is done for various temperatures, $U=1$ and a frequency grid of size $512\times512$. In Fig.~\ref{fig:sym_sus} we plot the RMSE as a function of the threshold quantile used for the truncation of the wavelet coefficients. Whereas we observe a constant and relatively low error for \emph{Haar} wavelets, in the case of the biorthogonal basis the RMSE is not only significantly larger but also gradually deteriorates upon increasing the threshold quantile. In order to obtain a better insight into this behaviour, we have explicitly plotted the reconstructed and symmetrized susceptibilities for $U=1, \beta=1000$ in Fig.~\ref{fig:sym_compare_bior} alongside the \emph{pixel-wise} absolute difference between the two. For the compression in this case we chose $l=4$ and $q=0.99$. In contrast to \emph{Haar} wavelets, where the plot of the absolute difference showed essentially nothing but noise (see Fig.~\ref{fig:sym_compare}), here we indeed observe distinct shapes emerging which are evidently due to the biorthogonal wavelets' inability to perfectly capture the symmetries of the generalized susceptibilities.

\end{appendices}

%
% For  figures use
%\begin{figure*}
%\vspace*{5cm}       % Give the correct figure height in cm
%\includegraphics{leer.eps}
%\end{figure*}
% or  this
%\begin{figure}
%\centering
%\resizebox{0.75\textwidth}{!}{%
%  \includegraphics{leer.eps}
%\vspace{5cm}       % Give the correct figure height in cm
%\end{figure}
%\begin{table}
%\centering
%\begin{tabular}{lll}
%\hline\noalign{\smallskip}
%\end{tabular}
% Or use
%\vspace*{5cm}  % with the correct table height
%\end{table}

%
% BibTeX users please use
% \bibliographystyle{}
% \bibliography{}
%\bibliographystyle{}
%\bibliography{refs, library}

\printbibliography
%\begin{thebibliography}{}
%Author, \textit{Book title} (Publisher, place year) page numbers
% etc
%\end{thebibliography}

\end{document}